\documentclass[11pt]{article}
\usepackage[dvips]{graphicx}
\def\Preprint{{\flushright{IFIC/01-14 \\ 
    FTUV/01-0323 \\ }}}

\newcommand{\beq}{\begin{equation}}
\newcommand{\eeq}{\end{equation}}

\newcommand{\lsim}{\stackrel{<}{_\sim}}
\newcommand{\gsim}{\stackrel{>}{_\sim}}

\begin{document}
\Preprint
\thispagestyle{empty}
\vspace{1cm}
\begin{center}
\begin{Large}
{\bf The NNLO $\tau^+\tau^-$ Production Cross Section Close to Threshold
} \\[1.75cm]
\end{Large}
{\large P. Ruiz-Femen\'\i a \ and \ A. Pich}\\[0.5cm]
{\it Departament de F\'\i sica Te\`orica, IFIC, Universitat de Val\`encia -
CSIC\\
 Apt. Correus 22085, E-46071 Val\`encia, Spain }\\[2cm]

\begin{abstract}
\noindent
The threshold behaviour of the
cross section $\sigma(e^+e^-\to\tau^+\tau^-)$ is analysed,
taking into account the known higher--order corrections.
At present, this observable can be determined to
next-to-next-to-leading order (NNLO) in a combined expansion in powers of
$\alpha_s$ and fermion velocities.
\end{abstract}
\end{center}
\vfill\eject

\pagenumbering{arabic}

\section{Introduction}

\hspace{0.5cm}
The Tau--Charm Factory, a high--luminosity ($\sim 10^{33}\;\mbox{cm}^{-2}\;
\mbox{s}^{-1}$) $e^+e^-$ collider with a centre--of--mass energy
near the $\tau^+\tau^-$ production threshold, has been proposed
\cite{KI:87,JO:87} as a powerful tool to perform high--precision studies
of the $\tau$ lepton, charm hadrons and the charmonium system
\cite{tcfSLAC,marbella}.
In recent years, this energy region has been only partially explored
by the Chinese BEBC machine ($\sim 10^{31}\;\mbox{cm}^{-2}\;\mbox{s}^{-1}$).
The possibility to operate the Cornell CESR collider
around the $\tau^+\tau^-$ threshold \cite{cornell}
has revived again the interest on Tau--Charm Factory physics \cite{PI:94}.

A precise understanding of the $e^+e^-\to\tau^+\tau^-$ production
cross section near threshold is clearly required. The accurate
experimental analysis of this observable could allow to improve the
present measurement \cite{BES} of the $\tau$ lepton mass.
The cross section $\sigma(e^+e^-\to\tau^+\tau^-)$ has already been 
analysed to ${\cal O} (\alpha^3)$ in refs.~\cite{voloshin,perrottet,SV:94},
including a resummation of the leading Coulomb corrections.

The recent development of non-relativistic effective
field theories of QED (NRQED) and QCD (NRQCD)  \cite{lepage}
has allowed an extensive investigation of the threshold production of
heavy flavours at $e^+e^-$ colliders. The threshold $b\bar b$ 
\cite{jamin,kuhn,bb}
and $t\bar t$ \cite{topprodsummary}
production cross sections have been computed to the
next-to-next-to-leading order (NNLO) in a combined expansion in powers of
$\alpha_s$ and the fermion velocities.
Making appropriate changes, those calculations can be easily applied to
the study of $\tau^+\tau^-$ production.

In this paper we will compile and analyse the known higher--order corrections
to the $\tau^+\tau^-$ production cross section. Although some ${\cal O} (\alpha^4)$
contributions have not been computed yet, the dominant NNLO corrections can
be already incorporated to the numerical predictions. One can then achieve
a theoretical precision better than 0.1\%.

The perturbative ${\cal O} (\alpha^3)$ and ${\cal O} (\alpha^4)$
contributions are discussed in section~\ref{sec:pert}.
Section~\ref{sec:NRQED} contains the relevant non-relativistic
corrections at low velocities, generating
${\cal O} (\alpha^n/v^m)$ effects.
The photon vacuum--polarization and the initial state
radiation contributions are accounted for in
sections~\ref{sec:vp} and \ref{sec:isr}, respectively.
In section~\ref{sec:ew}, electroweak corrections are shown to be 
negligible.
The numerical results for the $e^+e^-\to\tau^+\tau^-$ cross section and our
final conclussions are given in section~\ref{sec:numerics}. Some technical
details and detailed formulae are relegated to the appendices.

\section{The Perturbative Calculation up to ${\cal O} (\alpha^4)$}
\label{sec:pert}
\hspace{0.5cm}A NNLO analysis of a QED quantity, following
perturbation theory in the number of loops, implies that
contributions up to ${\cal O}(\alpha^4)$ should be taken into account. 
Let us review the
terms contributing to the total cross section of $\tau$ production in
$e^+e^-$ annihilation up to this order.  

At lowest order in QED, the $\tau$ leptons are produced by one-photon exchange
in the s-channel, and the total cross section formula reads
\begin{equation}
\sigma_{\mbox{\tiny $B$}} (e^+\,e^- \to \tau^+ \, \tau^-) ={{2 \pi \, \alpha^2} \over {3s}}\,
v\, (3-v^2)~~,
\label{s0}
\end{equation}
where $v=\sqrt{1-4 M^2/s}$ is the velocity of the final $\tau$ 
leptons in the center-of-mass 
frame of the $e^+ \, e^-$ pair and $M\equiv m_{\tau}$ is the $\tau$ mass.
$v$ is an adequate expansion parameter
for observables evaluated at energies close to the production threshold,
since its value goes to zero as we approach this point. This makes
$\sigma_{\mbox{\tiny $B$}}$
vanish in that limit, being the global factor $v$ in (\ref{s0}) of kinematic
origin. The quantum numbers of the $\tau^+ \, \tau^-$ pair are those of the
photon, $J^{PC}=1^{--}$, which corresponds to allowed $\tau^+ \, \tau^-$ states
$^3S_1$ and $^3D_1$ in spectroscopic notation $^{2S+1}L_J$.

Electromagnetic corrections of ${\cal O} (\alpha)$ to
$\sigma_{\mbox{\tiny $B$}}$ arise from the
interference between the tree level result and the following
1-loop amplitudes:
\begin{itemize}
\item[\it i)] ${\cal O} (\alpha)$ corrections to the $e^+e^-\gamma$ vertex,
\item[\it ii)] ${\cal O} (\alpha)$ corrections to the $\tau^+\tau^-\gamma$ vertex,
\item[\it iii)] vacuum polarization,
\item[\it iv)] box diagrams (2-photon production).
\end{itemize}

The contributions from {\it i)} and {\it ii)} are usually expressed in terms of
the Dirac and Pauli form factors at one loop \cite{barbi}. The corrections to
the  photon propagator {\it iii)} are divided into two pieces: the leptonic
contribution ($\ell=e,\mu,\tau$), which can be calculated perturbatively in QED,
and the hadronic contribution, where QCD corrections make a perturbative
estimate at low energies unreliable.
The hadronic vacuum polarization
can be related to the total cross section of hadron production by means of a
dispersion relation.
Finally, the interference of the tree--level amplitude with box diagrams
{\it iv)} does not
contribute to the total cross section, by virtue of Furry's Theorem.

Besides the above virtual radiative corrections, the cross section of
${\cal O} (\alpha^3)$ corresponding to the process of real photon emission,  
$e^+\,e^- \to \tau^+ \, \tau^- \,\gamma$\,, must be added. The Bremsstrahlung
photon can be emitted by the initial or final fermion lines, but there is no
contribution to the total cross section from the interference between both sets of
diagrams, again due to Furry's Theorem. We clearly see that there is no overlap
between initial and final state radiative corrections for the total cross section
up to ${\cal O} (\alpha^3)$. A compilation of analytical expressions for all the
terms mentioned above is found in Ref.~\cite{perrottet}.

Let us consider next ${\cal O} (\alpha^2)$ electromagnetic corrections to the
Born cross section.
They come from several sources:
\begin{itemize}
\item[$\bullet$] Interferences between the one-loop diagrams mentioned
previously. The
total cross section contributions from interferences between {\it i)}, {\it ii)}
and {\it iii)} with box diagrams are again zero. The first term involving
two-photon $\tau$ production comes from the square amplitude of the box diagrams.
\item[$\bullet$] Interferences between the Born term and the
following two-loop amplitudes:
the electron and the $\tau$ vertex two-loop corrections, contained in the
expressions of the electromagnetic form factors,
${\cal O} (\alpha)$ corrections to the
vacuum polarization, and three-photon production diagrams, for which only the
real part is needed.
\item[$\bullet$] The ${\cal O} (\alpha^4)$ Bremsstrahlung cross section, coming
from the interference between tree-level and one-loop diagrams with one
radiated photon, and from tree-level diagrams with two photons attached in
any of the fermion lines, corresponding to the process
$e^+\,e^- \to \tau^+ \, \tau^- \,\gamma \gamma$. It is no longer true, at this
order, that initial and final state real radiation could not interfere.
\end{itemize}

Recall that the spectral density $\mbox{Im} \Pi_{\mbox{\tiny em}}(s)$ built
from the electromagnetic
current of the $\tau$ leptons collects all final--state
interactions, including both virtual and real radiation, for single-photon
production, that is
\begin{equation}
\tilde{\sigma}(e^+e^-\to\gamma^*\to \tau^+ \, \tau^- )
\, = \,
\frac{48\pi^2\alpha^2}{3s}\,\mbox{Im}\Pi_{\mbox{\tiny em}}(s)
\,,
\label{spectral}
\end{equation}
where the tilde on $\sigma$ distinguishes it from the physical 
total cross section
which includes all kind of corrections. Relation (\ref{spectral}) results from a
direct application of the optical theorem, and is more commonly written
as the ratio
\begin{eqnarray}
R_{\mbox{\tiny $em$}}(s) & = &
\frac{\tilde{\sigma}(e^+e^-\to\gamma^*\to \ell^+\ell^-)}
{\sigma_{pt}}
\, = \,
12\pi \,\mbox{Im}\Pi_{\mbox{\tiny em}}(s)
\,,
\label{R(s)em}
\end{eqnarray}
i.e., normalizing $\tilde{\sigma}$ to the point cross section $\sigma_{pt}\, = \,
\frac{4\pi\alpha^2}{3s}$.
The ratio $R_{\mbox{\tiny $em$}}$ is well suited for
studying the non-relativistic dynamics of the $\tau^+ \, \tau^-$ pair,
as it fully contains 
the final--state interaction. Therefore, the threshold behaviour of the
total cross section will be ruled by the expansion of
$R_{\mbox{\tiny $em$}}$ at low velocities. The perturbative QED expression of 
$R_{\mbox{\tiny $em$}}$ is given in Appendix A up to NNLO in the combined expansion
in powers of $\alpha$ and $v$.

As long as we do not care about multiple photon production of $\tau$ leptons,
neither consider interference between initial and final state radiation,
it is possible to factorize the total cross section as an integration over
the product of separate pieces including initial, intermediate and
final state corrections:
\begin{equation}
\sigma(s)=\int^{s} \, F(s,w)\,\bigg|\frac{1}{1+e^2\Pi{\mbox{\tiny em}}(w)}
\bigg|^{2}\,
\tilde{\sigma}(w)\,dw \,.
\label{secradR}
\end{equation}
The radiation function $F(s,w)$ \cite{kuraev} describes initial state radiation,
including virtual corrections, and $\sqrt{s}$ is the total energy in the 
center-of-mass frame. The integration emerges to account for the effective energy loss
due to photon emission from the $e^+ \, e^-$ pair.
As previously mentioned, (\ref{secradR}) is an exact relation
for the total cross section only up to ${\cal O}(\alpha^3)$, but it
includes the largest corrections coming from the
emission of an arbitrary number of initial photons, which can sizeably
suppress the total cross section. The ${\cal O}(\alpha^4)$ contributions not
included in this analysis are those coming from two- and three-photon 
production diagrams, for which no velocity enhancement is expected in the 
threshold region and so represent pure ${\cal O}(\alpha^2)$ corrections 
$\sim 0.005\%$,
and the interferences between 2-photon Bremsstrahlung diagrams overlapping
initial and final state radiation. However, we shall argue in section~\ref{sec:NRQED}
that Bremsstrahlung contributions start at NNNLO in the
combined expansion in $\alpha$ and $v$, and so they are beyond the scope
of our analysis.

\section{Non-Relativistic Corrections: NRQED}
\label{sec:NRQED}

\hspace{0.5cm}We now focus on the behaviour of the total cross section in the
region just above the production threshold, where the small velocity of
the produced $\tau$ leptons is another
relevant parameter, besides $\alpha$. The final--state $\tau^+\tau^-$
interactions are encoded in the electromagnetic form factors. Written in
terms of $v$, their expressions
at one and two loops \cite{hoangver} show the existence of
${\cal O} \left(\frac{\alpha}{v}\right)$ and
${\cal O} \big(\frac{\alpha^2}{v^2}\big)$
power-like divergences
in the limit $v\to 0$. This is a general result for any number of loops:
diagrams with $n$ uncrossed photons exchanged between the produced leptons
generate singular terms proportional to $\left( \frac{\alpha}{v} \right)^n$,
known as Coulomb singularities, which lead to a breakdown of the QED
perturbative series in $\alpha$ when $v\to 0$. Resummation of such terms is
therefore mandatory, and it was done a long time ago \cite{sommer}, resulting in
the well-known Sommerfeld factor
\begin{equation}
|\Psi_{c,\mbox{\tiny E}}(0)|^2 \, = \, \frac {\alpha\pi/v}{1-\mbox{exp}(-
 \alpha\pi/v)}
\,,
\label{culfactor}
\end{equation}
multiplying the Born cross section~(\ref{s0}).
This factor corresponds to the wave function at the origin, solution
of the Schr\"odinger
equation, of two conjugate charged particles of mass $M$
interacting through a Coulomb potential for positive energies $E=Mv^2$.
The appearence of this factor in the cross section can be intuitively
understood, since the Coulomb interaction modifies the scattered wave function
of the lepton pair. 
The $1/v$ behaviour of
this factor makes the cross section at threshold finite.

We clearly see that a NNLO calculation of the cross section
in the kinematic region where
$\alpha \sim v$ has to account for all terms proportional
to $v\, (\alpha/v)^n\times[1;\alpha;v;\alpha^2;\alpha v;v^2]$ with
$n=1,2,\dots$ 
The leading divergences (i.e.
$\left( \frac{\alpha}{v} \right)^n$, $n > 1$) can be treated by using
well-known results from non-relativistic quantum mechanics, but a systematic
way to calculate higher-order corrections in this regime, such as
$\left( \frac{\alpha}{v} \right)^n\times [v ,v^2,\cdots]$,
seems to be
far from obvious, at least from the point of view of covariant perturbation
theory in the number of loops. An adequate description would come
from a simplified
theory which keeps the relevant physics at the scale $Mv \sim M\alpha$,
characteristic of the Coulomb interaction, allowing for a clear and systematic
identification of leading contributions. 

NRQED \cite{lepage} was designed precisely for this purpose. It is an effective
field theory of QED at low energies, applicable to fermions in non-relativistic
regimes, i.e. with typical momenta $p/M \sim v \ll 1$. Interactions contained in the
NRQED Lagrangian (\ref{NRQEDLagrangian}) have a definite velocity counting but
propagators and loop integrations can also generate powers of $v$. With appropriate
counting rules at hand, one can prove that all interactions between the non-relativistic
pair $\tau^+\tau^-$ can be described up to NNLO in terms of time-independent potentials
\cite{labelle}, derived from the low-energy Lagrangian.  
It can also be shown that the contributions to the total cross
section from diagrams with real photons emitted
from the produced heavy leptons begin at NNNLO~\footnote{This result can
be explicitly seen by going to the well-known expression for
$\sigma(e^+e^-\to \ell^+\ell^-\gamma)$ at tree level (see e.g.
\cite{berends}); the leading term is $\propto \alpha v^2$, i.e. NNNLO
compared to LO terms $\sim (\alpha/v)^n \sim {\cal O}(1)$.}.\par       
The key observable to study threshold effects in $\tau^+\tau^-$ production
is the 2-point function, 
$\Pi_{\mbox{\tiny em}}(s)$ calculated at NNLO.
Its fully covariant expression is written as the
time ordered correlator of two electromagnetic QED currents of the $\tau$
lepton $j^\mu = {\bar \tau}\gamma^\mu \tau$:
\begin{equation}
R_{\mbox{\tiny $em$}}(q^2) = \frac {4\pi}{q^2}\,\,\mbox{Im}\Bigl[ -i \int d^4x \, e^{iqx}\,
\langle 0| T\left( j^{\mu}(x) \,
j_{\mu}(0)^{\dagger} \right)|0\rangle \Bigr]
\,.
\label{Rcorrelator}
\end{equation} 
Inserting the effective low-energy expression for the QED current,
eq.~(\ref{corrienteNR}), 
in the last equation, one can arrive to the basic relation between the spectral density at
NNLO and the non-relativistic Green's functions \cite{hoangteubner}:
\begin{equation}
R^{\mbox{\tiny NNLO}}_{\mbox{\tiny $em$}}(q^2)  = 
\frac{6\,\pi}{M^2} \,\, \mbox{Im} \Big( C_1\,
G({\mbox{\boldmath $0$}},{\mbox{\boldmath $0$}};E)
\,-\frac{4E}{3M} \,
G_c({\mbox{\boldmath $0$}},{\mbox{\boldmath $0$}};E)
\Big)
\,,
\label{R(s)NNLOmain}
\end{equation}
with $C_1$ a short distance coefficient to be determined by matching full and effective
theory results. The details of this derivation are found
in Appendix B.

The Green's function $G$ obeys the Schr\"odinger equation corresponding 
to a two-body system
interacting through potentials derived from ${\cal{L}}_{\mbox{\tiny NRQED}}$
at NNLO, that means suppressed at most by $\alpha^2, \alpha/M$ or $1/M^2$, as
dictated by the counting rules. Such potentials have been calculated in the
literature \cite{fischler&billoire,peter,landau}, and in configuration space they
read 
\begin{eqnarray}
V_c(r) & = & -\,\frac{\alpha(\mu_s)}{r}\,
\bigg\{\, 1 +
\bigg(\frac{\alpha(\mu_s)}{4\,\pi}\bigg)\,\Big[\,
2\,\beta_1\,\ln(\tilde\mu\,r) + a_1
\,\Big]
\nonumber\\[2mm] & &
 +\, \bigg(\frac{\alpha(\mu_s)}{4\,\pi}\bigg)^2\,\Big[\,
\beta_1^2\,\Big(\,4\,\ln^2(\tilde\mu\,r)
      + \frac{\pi^2}{3}\,\Big)
+ 2\,\Big(2\,\beta_1\,a_1 + \beta_2\Big)\,\ln(\tilde\mu\,r)
+ a_2
\,\Big]
\,\bigg\}
\,,
\nonumber\\
\label{coulombpotential}
\\[2mm]
V_{\mbox{\tiny BF}}({\mbox{\boldmath $r$}}) & = &
\frac{\alpha(\mu_s)\,\pi}{M^2}\,
\,\delta^{(3)}({\mbox{\boldmath $r$}})
+ \frac{\alpha(\mu_s)}{2\,M^2 r}\,\Big[\,
{\mbox{\boldmath $\nabla$}}^2 + \frac{1}{r^2} {\mbox{\boldmath $r$}}\,
 ({\mbox{\boldmath $r$}} \,
{\mbox{\boldmath $\nabla$}} ) {\mbox{\boldmath $\nabla$}}
\,\Big]
\nonumber\\[2mm] & &
- \,\frac{\alpha(\mu_s)}{2\,M^2}\,
\left[\,
\frac{{\mbox{\boldmath $S$}}^2}{r^3}- \,
3\,\frac{\big({\mbox{\boldmath $S$}}\,{\mbox{\boldmath $r$}}\,\big)^2}{r^5}
\,-\frac{4\pi}{3}\big(2\,{\mbox{\boldmath $S$}}^2-3\big)\,\delta^{(3)}
({\mbox{\boldmath $r$}})
\,\right]+ \,\frac{3\,\alpha(\mu_s)}{2\,M^2\,r^3}\,{\mbox{\boldmath $L$}}\,
{\mbox{\boldmath $S$}}
\label{BFpotential}
\,,
\\[3mm]
V_{\mbox{\tiny An}}({\mbox{\boldmath $r$}}) & = &
\frac{\alpha(\mu_s)\,\pi}{M^2}\,{\mbox{\boldmath $S$}}^2
\,\delta^{(3)}({\mbox{\boldmath $r$}})
\label{potencialANI}
\,.
\\[3mm]
V_{\mbox{\tiny Ki}}({\mbox{\boldmath $r$}}) & = &
- \frac{{\mbox{\boldmath $\nabla$}}^4}{4M^3}
\label{kinpote}
\,.
\end{eqnarray}
Here $\alpha(\mu_s)$ denotes the electromagnetic coupling constant 
renormalized in the
$\overline{MS}$ scheme at the scale $\mu_s \equiv \mu_{\mbox{\tiny $soft$}}$.
The latter is the
renormalization scale set for the ${\cal O}(\alpha)$ and
${\cal O}(\alpha^2)$ corrections to the Coulomb
potential (\ref{coulombpotential}), as determined 
in \cite{fischler&billoire} and \cite{peter},
respectively. Note that these corrections involve ultraviolet divergent
light fermion loops
($m_f\ll M$), which
cannot be accurately described within NRQED. The scale $\tilde{\mu}$ is equal to
$\mu_{\mbox{\tiny $soft$}}\, e^{\gamma_{\mbox{\tiny E}}}$, with
$\gamma_{\mbox{\tiny E}}$ the Euler constant, and the rest of coefficients in
(\ref{coulombpotential}) take the values
\begin{eqnarray}
\beta_1 & = & - \frac{4}{3}\,n_{\ell}
\,,
\hspace{1.0cm}
\beta_2  =  - 4\,n_{\ell}
\,,
\label{betacoeff}\\[2mm]
a_1 & = & - \frac{20}{9}\,n_{\ell}
\,,
\hspace{1.0cm}
a_2  =
-\bigg(\,\frac{55}{3}-16\,\zeta_3\,\bigg)\,n_{\ell}
+\bigg(\,\frac{20}{9}\,n_{\ell}\,\bigg)^2
\,.
\label{ctesestruc}
\end{eqnarray}  
The constants $\beta_1$ and $\beta_2$ are the one- and two-loop coefficients
of the QED beta function in the $\overline{MS}$ scheme defined as
\begin{equation}
\frac{d\ln\alpha}{d\ln\mu^2}=\beta(\alpha)=
\beta_1\frac{\alpha}{4\pi}+\beta_2
\bigg(\frac{\alpha}{4\pi}\bigg)^2+\cdots
\label{beta-def}
\end{equation}
The number of active lepton flavors $n_{\ell}$ would be equal to two for interacting
$\tau$'s. If quark loops are included we should substitute
$n_{\ell}\to n_f\equiv(n_{\ell}+N_c \sum_q Q_q^2)$, being $Q_q$ the electromagnetic
charge of the quark $q$ (with mass lower than $M$). 

The Breit-Fermi potential $V_{\mbox{\tiny BF}}$ 
(see e.g. \cite{landau}) has been written in terms of the total spin 
${\mbox{\boldmath $S$}}$ and angular momentum ${\mbox{\boldmath $L$}}$ of
the lepton pair. At NNLO, the heavy leptons are only produced in triplet 
S-wave states, so we just need to consider the corresponding projection of the 
$V_{\mbox{\tiny BF}}$ potential,
(i.e. make ${\mbox{\boldmath $S$}^2}=2$ and ${\mbox{\boldmath $L$}}=0$
in (\ref{BFpotential})). $V_{\mbox{\tiny An}}$ is a NNLO piece
derived from the first contact
term written in ${\cal{L}}_{\mbox{\tiny NRQED}}$, eq.~(\ref{NRQEDLagrangian}),
which reproduces the QED tree level
s-channel diagram for the process $\ell^+\ell^-\to \ell^+\ell^-$. In QCD this
diagram connects $q\bar{q}$ color-octet states, so this piece is not
present in recent papers devoted to threshold electromagnetic quark
production, where $q\bar{q}$ pairs can only be produced in color-singlet states.
Finally, the
term (\ref{kinpote}) is the first relativistic correction to the kinetic energy.

The Green's function at NNLO will therefore satisfy the Schr\"odinger
equation~\footnote{Note that the Green's function built from the NNLO potentials also
resums higher order contributions, like those diagrams with the insertion of
more than one NNLO potential term.}
\begin{equation}
\bigg( -\frac{{\mbox{\boldmath $\nabla$}}^2}{M} -
\frac{{\mbox{\boldmath $\nabla$}}^4}{4M^3} +
V_{c}({\mbox{\boldmath $r$}}) + V_{\mbox{\tiny BF}}({\mbox{\boldmath $r$}})
+ V_{\mbox{\tiny An}}({\mbox{\boldmath $r$}})
-E\,\bigg)\,G({\mbox{\boldmath $r$}},{\mbox{\boldmath $r$}^\prime},E)
\, = \,
\delta^{(3)}({\mbox{\boldmath $r$}}-{\mbox{\boldmath $r$}^\prime})
\,.
\label{Schrodingerfull}
\end{equation}
A solution of eq.~(\ref{Schrodingerfull}) must rely on numerical or perturbative
techniques. In the QED case, a significant difference between both approaches is
not expected, being $\alpha$ such a small parameter~\footnote{Although
for heavy quarks the numerical solution of the
Schr\"odinger equation has been shown to have more stable NLO and NNLO
corrections,
we should note that higher-order terms not under control are being resummed,
some of which are cutoff-dependent \cite{topprodsummary}.}.
Consequently we will follow the perturbative approach, using recent
results by Hoang,
Penin and others \cite{kuhn,hoangteubner,peninphyslett}, who calculated the NLO and NNLO corrections to the Green's
function analytically, via
Rayleigh-Schr\"odinger time-independent perturbation theory around the
known LO Coulomb Green's function:
\begin{eqnarray}
G({\mbox{\boldmath $x$}},{\mbox{\boldmath $y$}};E)\,=\,
{G}_c({\mbox{\boldmath $x$}},{\mbox{\boldmath $y$}};E)\,+\,
\delta G({\mbox{\boldmath $x$}},{\mbox{\boldmath $y$}};E)\,,
\nonumber
\label{Gcorrec}
\end{eqnarray}
\begin{eqnarray}
\delta G({\mbox{\boldmath $x$}},{\mbox{\boldmath $y$}};E)\,=\,
-\int d^3 \mbox{\boldmath $z$}\,
{G}_c({\mbox{\boldmath $x$}},{\mbox{\boldmath $z$}};E)\,(H-H_0)\,
{G}_c({\mbox{\boldmath $z$}},{\mbox{\boldmath $y$}};E)\,+\cdots\,
\nonumber
\end{eqnarray}
\vspace*{-0.5cm}
\begin{eqnarray}
= -\int d^3 \mbox{\boldmath $z$}\,
{G}_c({\mbox{\boldmath $x$}},{\mbox{\boldmath $z$}};E)\,
\bigg( -\frac{{\mbox{\boldmath $\nabla$}}^4}{4M^3} +
V_{\mbox{\tiny BF}}({\mbox{\boldmath $z$}}) +
V_{\mbox{\tiny An}}({\mbox{\boldmath $z$}}) +
V_{c}^{\tiny (1)}({\mbox{\boldmath $z$}}) +
V_{c}^{\tiny (2)}({\mbox{\boldmath $z$}})\bigg)\,
{G}_c({\mbox{\boldmath $z$}},{\mbox{\boldmath $y$}};E)\,+\cdots
\nonumber
\end{eqnarray}
\vspace{-0.1cm}
\begin{equation}
\hspace*{2.2cm}\, =\,
\delta_{\mbox{\tiny Ki,BF}}G+\delta_{\mbox{\tiny An}}G
+\delta^{\mbox{\tiny NLO}}_{\tiny 1} G+\delta_{\tiny 2} G+
\delta^{\mbox{\tiny NNLO}}_{\tiny 1}G+\,\cdots
\,.
\label{perturG}
\end{equation}
Here $H_0=-{\mbox{\boldmath $\nabla$}}^2/M+
V_{c}^{\mbox{\tiny LO}}(r)$ is the pure Coulomb Hamiltonian. We refer the 
reader to Appendix~C for complete expressions of $G_c$ and the different
$\delta G$'s, as calculated in the literature, and for a full discussion 
about the regularization procedure.
Let us just quote here that the Sommerfeld factor (\ref{culfactor}),
which appears in the LO cross
section can be easily recovered from the basic relation (\ref{R(s)NNLOmain}), if
one reminds the spectral representation of the Green's function
\begin{equation}
G({\mbox{\boldmath $r$}},{\mbox{\boldmath $r$}^\prime};E)  =  
\sum_{n} \,
\frac{\Psi_{n}({\mbox{\boldmath $r$}}) \,
\Psi_{n}^{*}({\mbox{\boldmath $r$}^\prime})}{E_n-E-i\epsilon}\,+
\,\int\frac{d^3{\mbox{\boldmath $k$}}}{(2\pi)^3} \,
\frac{\Psi_{k}({\mbox{\boldmath $r$}}) \,
\Psi_{k}^{*}({\mbox{\boldmath $r$}^\prime})}
{E_k-E-i\epsilon}
\,,
\label{autofunc}
\end{equation}
with $\Psi_n({\mbox{\boldmath $r$}})$ the bound state's wave functions 
($E_n<0$),
and $\Psi_{k}({\mbox{\boldmath $r$}})$ corresponding to eigenfunctions of $H$ 
with $E_k={\mbox{\boldmath $k$}}^2/M>0$. The LO spectral density is proportional
to the imaginary part of the Coulomb Green's function, and so, from (\ref{autofunc}),
proportional to $|\Psi_{c,\mbox{\tiny E}}(0)|^2$, i.e. to the solution at the origin of 
the Schr\"odinger equation with the LO Coulomb potential.     

Finally, the short distance coefficient $C_1$ must be fixed.      
The ``direct matching procedure'' \cite{hoangvac} allows 
a straightforward determination of $C_1$, by
comparing the NNLO non-relativistic expression (\ref{R(s)NNLOmain}) 
with the result (\ref{R(s)NNLO QED}) for
$R_{\mbox{\tiny $em$}}$,
calculated in full QED keeping terms up to ${\cal O}(\alpha^2)$
and NNLO in the velocity expansion. 
The short distance coefficient $C_1$
is then expressed as a perturbative series in
$\alpha(\mu_{\mbox{\tiny $hard$}})$
\newpage
\begin{equation}
C_1(M,\mu_{\mbox{\tiny $hard$}},\mu_{\mbox{\tiny $fac$}})\,=\,1+\left(
 \frac{\alpha(\mu_{\mbox{\tiny $hard$}})}{\pi}\right)\,c_1^{(1)}+
\left(\frac{\alpha(\mu_{\mbox{\tiny $hard$}})}{\pi}\right)^2\,c_1^{(2)}(\mu_{\mbox{\tiny $hard$}},
\mu_{\mbox{\tiny $fac$}})+\dots
\,,
\label{C1}
\end{equation}
where we have anticipated that $c_1^{(1)}$ does not depend on any scale. The
renormalization point $\mu_{\mbox{\tiny $hard$}}$, chosen for
$\alpha_{\mbox{\tiny $\overline{MS}$ }}$
in the short distance coefficient, needs not be equal to that governing the
perturbative expansions of the correlators, $\mu_{\mbox{\tiny $soft$}}$, which 
only contains
long-distance physics~\footnote{Differences are relevant when NNLO corrections
are considered.}. The result of the matching reads \cite{hoangteubner}
\begin{eqnarray}
c_1^{(1)}&=&-4
\nonumber\\[2mm]
c_1^{(2)}&=&\pi^2\,\bigg[\,\kappa -\frac{4}{3\pi^2}n_f\,
\ln\frac{M^2}{\mu_{\mbox{\tiny $hard$}}^2}
-\frac{1}{6}\ln\frac{M^2}{\mu_{\mbox{\tiny $fac$}}^2}
\,\bigg]\,,
\label{c1s}
\end{eqnarray}
with
\begin{equation}
\kappa  =
\bigg[\, \frac{1}{\pi^2}\,\bigg(\,
\frac{39}{4}-\zeta_3 \,\bigg) +
\frac{4}{3} \ln 2 - \frac{35}{18}
\,\bigg] 
+
\bigg[\,
\frac{4}{9}\,\bigg(\, \frac{11}{\pi^2} - 1\,\bigg)
\,\bigg] +
n_f\,\bigg[\, \frac{11}{9\,\pi^2} \,\bigg]
\,.
\label{kappadef}
\end{equation}   
The factorization scale $\mu_{\mbox{\tiny $fac$}}$ is introduced to separate long and short distance
contributions in the process of regularization (see Appendix C
for details).

\section{Vacuum Polarization}
\label{sec:vp}

\hspace*{0.5cm}We now turn over intermediate state corrections
in formula (\ref{secradR}).
For a complete NNLO description of $\sigma(e^+e^- \to \tau^+\tau^-)$,
two-loop corrections to the photon propagator should be included.
Despite having calculated the final state observable
$R_{\mbox{\tiny $em$}}$ in the $\overline{MS}$ scheme, we can exploit the fact
that the piece $e^2/\,[1+e^2\,\Pi_{\mbox{\tiny em}}(s)]$ is a renormalization
group invariant, and so 
evaluate these set of corrections in the $on-shell$ scheme, where decoupling of
heavy fermions is naturally implemented. The $on-shell$ renormalized vacuum
polarization function is defined as
\begin{equation}
\Pi_{\mbox{\tiny em}}^{\mbox{\tiny $ren$}}(q^2)\,=\,
\Pi_{\mbox{\tiny em}}(q^2)-\Pi_{\mbox{\tiny em}}(0)
\,.
\label{onshell}
\end{equation}

The light lepton
contributions to the vacuum polarization are the standard
1- and 2-loop perturbative expressions \cite{kallen-sabry}:
\begin{equation}
e^2\,\Pi_{\mbox{\tiny $e,\mu$}}(q^2)\,=\,
\bigg(\frac{\alpha}{\pi}\bigg)\Pi^{(1)}(q^2)+
\bigg(\frac{\alpha}{\pi}\bigg)^2\Pi^{(2)}(q^2)+
{\cal O}(\alpha^3)
\,,
\label{polaQED}
\end{equation}
with
\begin{eqnarray}
\Pi^{(1)}(q^2) & = & \sum_{i=e,\mu}\,\frac{1}{3}\bigg[\frac{5}{3}-
\ln\Big(-\frac{q^2}{m_i^2}\Big)+\frac{6m_i^2}{q^2}+
{\cal O}\Big(\frac{m_i^4}{q^4}\Big)\bigg]\,,
\label{pol-light-lep1loop}\\[2mm]
\Pi^{(2)}(q^2) & = & \sum_{i=e,\mu}\,\frac{1}{4}\bigg[\frac{5}{6}-4\zeta_3-
\ln\Big(-\frac{q^2}{m_i^2}\Big)-
12\frac{m_i^2}{q^2}\ln\Big(-\frac{q^2}{m_i^2}\Big)+
{\cal O}\Big(\frac{m_i^4}{q^4}\Big)\bigg]
\,,
\label{pol-light-lep2loop}
\end{eqnarray}
where we have only retained the relevant terms in the limit
$m_{\ell}^2\ll q^2$ ($m_\ell$ are the pole light-lepton masses). 
For the $\tau$ contribution in the threshold vicinity $q^2 \gsim 4M^2$, 
resummation
of singular terms in the limit $v\to 0$ is mandatory. Under the assumption
$\alpha \sim v$, it is clear that we need to know NLO contributions to
$\Pi_{\tau}(q^2)$, which means retaining uniquely $G_c$ and
$\delta_ 1^{\mbox{\tiny NLO}}G$ in (\ref{perturG}), but performing the direct
matching not only for the imaginary part but also for the real part (up to
${\cal O}(\alpha)$): 
\begin{equation}
e^2\Pi^{\mbox{\tiny NLO}}_{\tau}(q^2)=
 \frac{2\pi\alpha}{M^2}\,C_1\bigg(\,
{G}_c({\mbox{\boldmath $0$}},{\mbox{\boldmath $0$}};E)+
\delta^{\mbox{\tiny NLO}}_1
G({\mbox{\boldmath $0$}},{\mbox{\boldmath $0$}};E)\bigg)+
\alpha \,h_1+\alpha^2 h_2\,.
\label{piNLO}
\end{equation}
The one-loop coefficient $C_1$ was already obtained in (\ref{c1s}),
and $h_1,h_2$ are fixed by demanding equality between
$\mbox{Re}\, \Pi_{\tau}$ calculated in full QED and expression (\ref{piNLO}). 
We get
\begin{eqnarray}
h_1&=&\frac{8}{9\pi}\,,
\nonumber\\[2mm]
h_2&=&\frac{1}{4\,{{\pi }^2}}\,\left( 3 - \frac{21}{2}\,\zeta_3 \right)
  + \frac{11}{32} -  \frac{3}{4}\,\ln 2 +\frac{1}{2}\,\ln
  \frac{M}{\mu_{\mbox{\tiny $fac$}}}\,.
\label{h1h2}
\end{eqnarray}

In the hadronic sector, a perturbative estimate of the vacuum
polarization in terms of free quarks is unreliable since strong interactions
at low energies become non-perturbative. An alternative approach consists in relating
the hadronic vacuum polarization with the total cross section
$\sigma(e^+e^-\to\gamma^*\to had)$, by using unitarity and the analyticity
of $\Pi_{\mbox{\tiny had}}(s)$:
\begin{eqnarray}
\Pi_{\mbox{\tiny had}}(s)&=&\frac{s}{\pi}\int^{\infty}_{4m_{\pi}^2}dt\,\,
\frac{\mbox{Im}\,\Pi_{\mbox{\tiny had}}(t)}{t(t-s-i\epsilon)}
\nonumber\\
&=& \frac{s^2}{16\pi^3\alpha^2}\,
\int^{\infty}_{4m_{\pi}^2}dt\,\,
\frac{\sigma(e^+e^-\to had)}{t(t-s-i\epsilon)}\,.
\label{disp-rel}
\end{eqnarray}
Usually, $\sigma(e^+e^-\to\gamma^*\to had)$ is conveniently parameterized and
the unknown parameters fitted to experimental measurements or else related to
phenomenological constants. In this paper we will  
make use of a parameter-free
formula for $\sigma(e^+e^-\to\gamma^*\to had)$ in the low-energy region,
where the non-perturbative effects are more important, 
and the perturbative result
for the high energy part. Below 1~GeV, the electromagnetic
production of hadrons is dominated by the $\rho$ resonance ($J^{PC}=1^{--}$)
and its decay to two charged pions. The photon mediated $\pi^+\pi^-$
production cross section at a center-of-mass energy $\sqrt{s}$ is
written as
\begin{equation}
\sigma(e^+e^-\to \pi^+\pi^-)=\frac{\pi\alpha^2}{3s}
\bigg(1-\frac{4m_{\pi}^2}{s}\bigg)^{3/2}\arrowvert\,F(s)\,\arrowvert^2
\,,
\label{seccPiones}
\end{equation}
with $F(s)$ being the pion electromagnetic form factor defined as
$$
\langle \pi^+\pi^-| j^{\mu} |0{\rangle}=F(s)\,
(p_{\pi^-}-p_{\pi^+})^{\mu}\,.
$$
In the isospin limit, only the $I=1$ part of the quark electromagnetic
current 
$j^{\mu}=Q_u\,\bar{u}\gamma^{\mu}u+Q_d\,\bar{d}\gamma^{\mu}d$
survives.
An analytic expression for the pion
isovectorial form factor was obtained in Ref.~\cite{guerrero} using Resonance
Chiral Theory \cite{RChiT} and restrictions
imposed by analyticity and unitarity. The so-obtained $F(s)$, which
provides an excellent description of experimental data up to energies
of the order
of 1~GeV, reads:
\begin{eqnarray}
F(s) &\!\!\! =&\!\!\!
\frac{M^2_{\rho}}{M^2_\rho-s-iM_\rho \Gamma_\rho (s)}
\exp \Biggl\{
\frac{-s}{96\pi^2 f_\pi^2}\hbox{Re}\,A(m_\pi^2/s,m_\pi^2/M_\rho^2)\Biggr\}
\,,
\label{F(s)}
\end{eqnarray}
where $\Gamma_\rho (s)$ is the {\it off-shell} width of the $\rho$ meson
\cite{jorge},
\begin{eqnarray}
\Gamma_\rho (s) & = & {M_\rho\, s\over 96\pi f_\pi^2}
\theta(s-4m_\pi^2)\,\sigma_{\pi}^3 \nonumber\\[2mm]
&=& -{M_\rho\, s\over 96\pi^2 f_\pi^2} \,
\hbox{Im}\left[ A(m_\pi^2/s,m_\pi^2/M_\rho^2)\right]
\,,
\label{rowidth}
\end{eqnarray}
and
\begin{equation}
A(m_{\pi}^2/s,m_{\pi}^2/M_{\rho}^2) \, =\,
\ln{\left( m^2_{\pi}/M_{\rho}^2\right)} + {8 m^2_{\pi}\over s} -
\frac{5}{3}  + \sigma_{\pi}^3 \,\ln{\left(\frac{\sigma_{\pi}+1}
{\sigma_{\pi}-1}\right)}
\,,
\nonumber
\label{Afunc}
\end{equation}
$$
\sigma_{\pi}\equiv \sqrt{1-4m_{\pi}^2/s}\,.
$$
Formula (\ref{seccPiones}) will be integrated in (\ref{disp-rel}) up to
an upper bound $s_{\rho} \sim $ 1~GeV$^2$. For the integration region above
$s_{\rho}$, we use the perturbative results of $\mbox{Im}
 \Pi_{\mbox{\tiny had}}$:
\begin{equation}
e^2\mbox{Im}\,\Pi_{u,d,s}(s)=\sum_{q=u,d,s}N_c\,Q_q^2\frac{\alpha}{3}
\left[1+\frac{\alpha_s}{\pi}\right]\,,
\label{Pi uds}
\end{equation}
for light quarks, in the zero mass limit, and
\begin{eqnarray}
e^2\,\mbox{Im}\,\Pi_{c,b}(s)=\sum_{q=c,b}\theta (s-4m_q^2)
N_c\,Q_q^2\frac{\alpha}{3}\!\!\!\!&\Bigg[&\!\!\!\!\Big(1+\frac{2m_q^2}{s}\Big)
\,\sqrt{1-\frac{4m_q^2}{s}}
\nonumber\\
\!\!\!\!&+&\!\!\!\!\frac{\alpha_s}{\pi}\,C_F
\bigg(\frac{3}{4}+9\frac{m_q^2}{s}
+\frac{m_q^4}{s^2}\Big(\frac{5}{2}-18\ln\frac{m_q^2}{s}\Big)\bigg)\,\Bigg]
\,,
\label{Pi cb}
\end{eqnarray}
for the $b$ and $c$ quarks~\footnote{At the energy scales of $\tau$ production
the $b$ quark has not been considered in the particle content of the effective 
theory, but we will include it when running $\alpha$ to $s=M_Z^2$. 
The contribution of the top quark to 
(\ref{disp-rel}) starts at $\sqrt{s}\simeq 350$~GeV, so it is highly suppressed
by the $t^2$ factor in the denominator.}. In both (\ref{Pi uds}) and
(\ref{Pi cb}) the first QCD loop correction to the quark vacuum polarization
has been added,
with $\alpha_s$ the strong
coupling constant. This simplified description is good enough to achieve an 
accuracy better than 0.1$\%$ for the $e^+e^- \to \tau^+\tau^-$ cross section.

As a test of our method to calculate the hadronic vacuum
polarization, we have computed its contribution to the running of $\alpha$ at the
scale $\sqrt{s}=M_Z$, and compared it with the results of recent analyses
devoted to this subject \cite{jegerlehner,davier}. In the {\it on-shell} scheme 
the evolution of the electromagnetic
coupling constant due to hadron polarization is commonly defined as
$$
\alpha(s)=\frac{\alpha}{1-\Delta_{\mbox{\tiny had}}\alpha(s)}
$$
with
$$
\Delta_{\mbox{\tiny had}}\alpha(s)=-4\pi\alpha\,\mbox{Re}
\left[\Pi_{\mbox{\tiny had}}(s)-
\Pi_{\mbox{\tiny had}}(0)\right]\,.
$$ 
At the scale $M_Z$ we get
$\Delta_{\mbox{\tiny had}}\alpha(M_Z^2)\times 10^{4}=268$,
to be compared with the values 
$\Delta_{\mbox{\tiny had}}\alpha(M_Z^2)\times 10^{4}=280 \pm 7$
and $\Delta_{\mbox{\tiny had}}\alpha(M_Z^2)\times 10^{4}=276.3\pm 1.6$, 
obtained in 
\cite{jegerlehner} and \cite{davier} respectively. 
Our simple estimate only deviates by 4$\%$ and 3$\%$ respectively, from those
analyses. Considering
that $\Pi_{\mbox{\tiny had}}$ modifies
$\sigma(e^+e^-\to \tau^+\tau^-)$ near threshold by roughly $1\%$, 
our result has a global
uncertainty smaller than $0.1\%$ for the total cross section. 

Let us just mention that the
theoretical description of the vector form factor of the pion has been 
improved in a recent paper \cite{Portoles} using a model-independent
parameterization which can fairly reproduce experimental data coming from 
$e^+e^-\to \pi^+\pi^-$ up to higher energies, 
$\sqrt{s} \lsim 1.2$~GeV. With
such results, we would gain knowledge on the hadronic contribution to vacuum
polarization, but its numerical effect on our final estimate would not be 
relevant,
considering the important features of the hadronic spectrum we
are leaving out by using naive QCD perturbation theory from
$\sqrt{s}\sim 1.2$~GeV upwards.

\section{Initial State Radiation}
\label{sec:isr}

\hspace*{0.5cm}In this section we collect the radiative corrections to
single-photon
annihilation of the initial $e^+ e^-$ pair. These include both virtual and
real photon radiation, all of which are needed at ${\cal O}(\alpha^2)$ in a 
formal NNLO analysis of $\sigma(e^+e^-\to \tau^+\tau^-)$. However, for the
emission of soft photons (i.e. photons whose energy do not exceed an
experimental resolution $\Delta E \ll \sqrt{s}$), it is a well-known feature
that
the expansion parameter is not $\alpha$ but $(\alpha/\pi)\log (s/m_e^2)\log
(E/\Delta E)$, which may be quite large, making necessary to retain
all terms of the expansion with respect to it. It is possible to perform
such resummation by using an approach based on the Structure Functions formalism
\cite{kuraev}. In this technique, the effect of initial state radiation is
accounted for by convoluting the cross section without initial radiative
corrections with Structure Functions for electrons and positrons, in analogy
with a Drell-Yan process in QCD. In the leading logarithmic approximation
(i.e. when only terms containing a factor $L\equiv\log (s/m_e^2)$ with each
power of $\alpha$ are retained) this formalism allows to represent the cross
section in the form (\ref{secradR}):
\begin{equation}
\sigma(s)=\int^{\frac{2\Delta E}{\sqrt{s}}}_0 dx \,F(x,s)\,
\bigg|\frac{1}{1+e^2\Pi(s^\prime)}\bigg|^{2}\,
\tilde{\sigma}(s^\prime)\,,
\label{secradR2}
\end{equation} 
with the ``available'' center-of-mass energy after Bremsstrahlung loss
defined as $s^\prime=s(1-x)$, and the radiation function
\begin{eqnarray}
F(x,s)\!\!&=&\!\!\beta x^{\beta-1}\bigg[1+\frac{\alpha}{\pi}
\bigg(\frac{\pi^2}{3}-
\frac{1}{2}\bigg)+\frac{3}{4}\beta-\frac{\beta^2}{24}\bigg(\frac{1}{3}L+
2\pi^2-\frac{37}{4}\bigg)\bigg]-\beta\bigg(1-\frac{1}{2}x\bigg)
\nonumber\\[2mm]
\!\!&+&\!\!\frac{1}{8}\beta^2\bigg[4(2-x)\ln\frac{1}{x}
-\frac{(1+3(1-x)^2)}{x}\ln(1-x)-6+x\bigg]\,,
\label{radfunc}
\end{eqnarray}
$$
\beta=\frac{2\alpha}{\pi}(L-1)\,.
$$
The total cross section, at the kinetic energy above threshold 
$E=\sqrt{s}-2M$, is evaluated by convoluting 
the photon-mediated cross section of $\tau^+\tau^-$ production without initial
radiative corrections with a weight function $F$ describing such radiation effects,
from an energy $E$ down to $E^\prime \simeq E-\Delta E$. The function $F(x,s)$ becomes
larger as $x\to 0$, i.e. for $E^\prime \lsim E$, and it strongly decreases as the
$x$ variable grows.
   
Besides the leading $(\alpha/\pi)^n L^n$ terms,
expression (\ref{radfunc}) also includes
all ${\cal O} (\alpha)$ terms exactly. The analysis of higher-order 
terms, not included in
$F(x,s)$, is showed in \cite{kuraev} to go beyond $0.1\%$ accuracy for the
interval of energies $0.2$~GeV$<\sqrt{s}<10$~GeV.  
We shall use eq. (\ref{radfunc}) to evaluate initial state corrections
to the total cross section. 

\section{Electroweak Corrections}
\label{sec:ew}

\hspace*{0.5cm}The small corrections arising from $\tau$ production through
a $Z$ boson can be easily incorporated in our
basic formula (\ref{secradR2}). Electroweak production of heavy quarks
including threshold effects
has already been studied in previous papers \cite{penin2,hoang-teubner2}. The
trivial part comes from the vector couplings of the $Z$ current, which just add
a term proportional to $R_{\mbox{\tiny $em$}}(s)$ to the total cross section:
\begin{equation}
\tilde{\sigma}^{\gamma^*\!,Z^*_{\tiny vec}}(s)
 = \sigma_{pt}\left[1-2\frac{s}{s-M_z^2}\,v_e\, v_{\tau}+
\left(\frac{s}{s-M_z^2}\right)^2\left[v_e^2 +
a_e^2\right]v_{\tau}^2\right]R_{\mbox{\tiny $em$}}(s)
\,,
\label{R(s)em+Z}
\end{equation}
where $v_{\ell}$ and $a_{\ell}$ are the neutral-current couplings of charged
leptons,
\begin{eqnarray}
v_{e,\mu,\tau} &=& \frac{-1+4\sin^2 \theta_W}{4\sin \theta_W \cos \theta_W}\,,
\\[2mm]
a_{e,\mu,\tau} &=& \frac{-1}{4\sin \theta_W \cos \theta_W}
\,.
\label{neutralcouplings}
\end{eqnarray} 
At the $\tau^+\tau^-$ threshold, electroweak corrections are at least
suppressed by terms of
${\cal O}(8m_{\tau}^2/M_Z^2) \sim 3\cdot 10^{-3}$ with respect to photon
mediated production. Due to the further suppression induced by the couplings 
$v_e=v_\tau \sim 0.05$, these electroweak
corrections represent a contribution below 0.0008\% to the total cross section,
and therefore, they will not be considered for our purposes. 

The non-trivial
part of the electroweak corrections comes from the axial couplings of the $Z$ boson
with the non-relativistic final state fermions. For such contributions
one needs to expand the QCD
axial-vector current in terms of proper NRQED currents and then to construct
the corresponding non-relativistic correlator, which is already a NNLO
contribution describing the $\tau^+\tau^-$ system in a P-wave triplet state 
\cite{penin2,hoang-teubner2}. However it is suppressed by
${\cal O}(16m_{\tau}^4/M_Z^4)$, so fully negligible in our analysis.

\section{Final Results for $\sigma(e^+e^- \to \tau^+\tau^-)$}
\label{sec:numerics}

\begin{figure}[!tb]
\includegraphics[angle=0,width=0.48\textwidth]{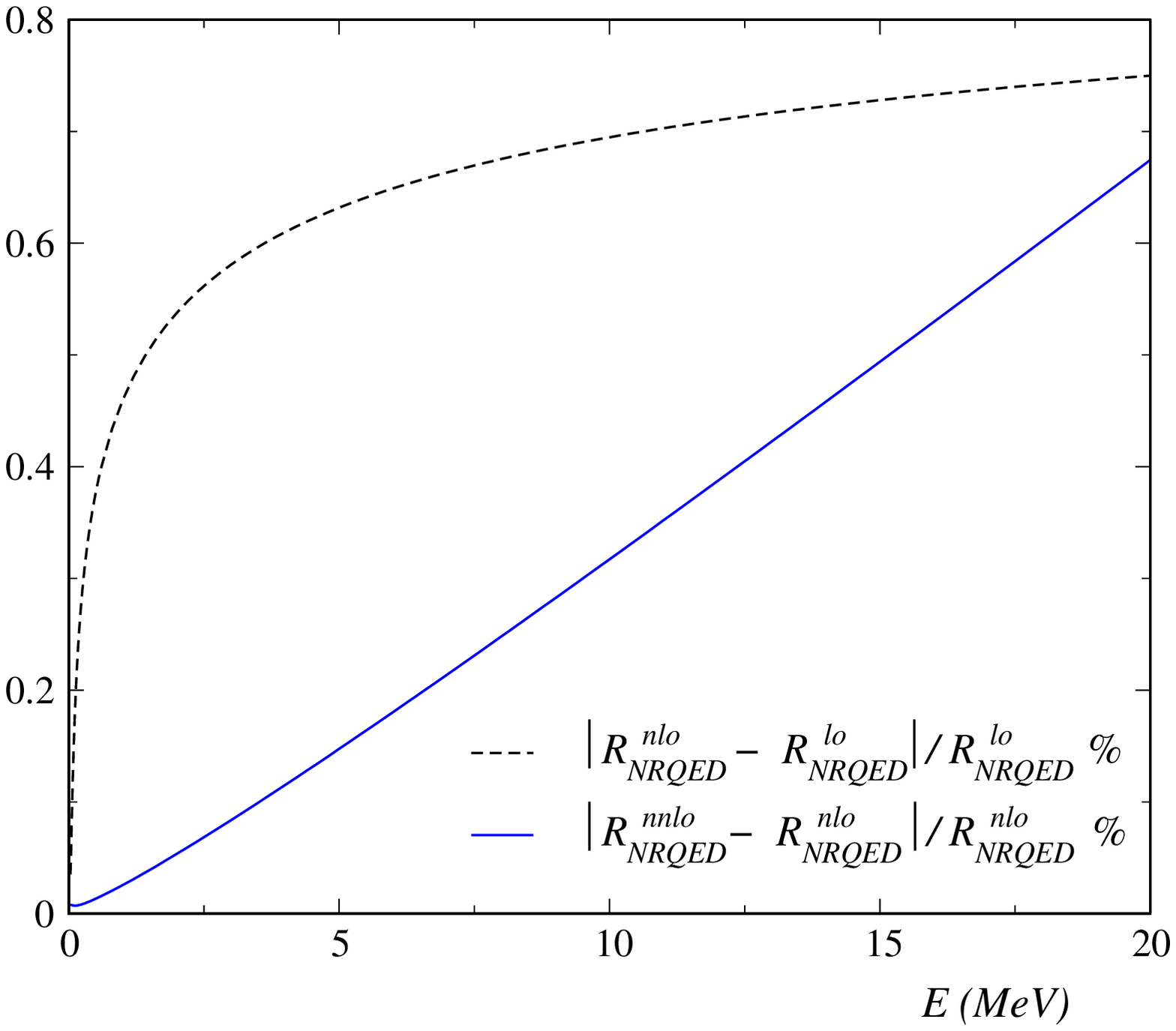}
\vspace*{-0.cm}
\hspace*{0.5cm}
\includegraphics[angle=0,width=0.48\textwidth]{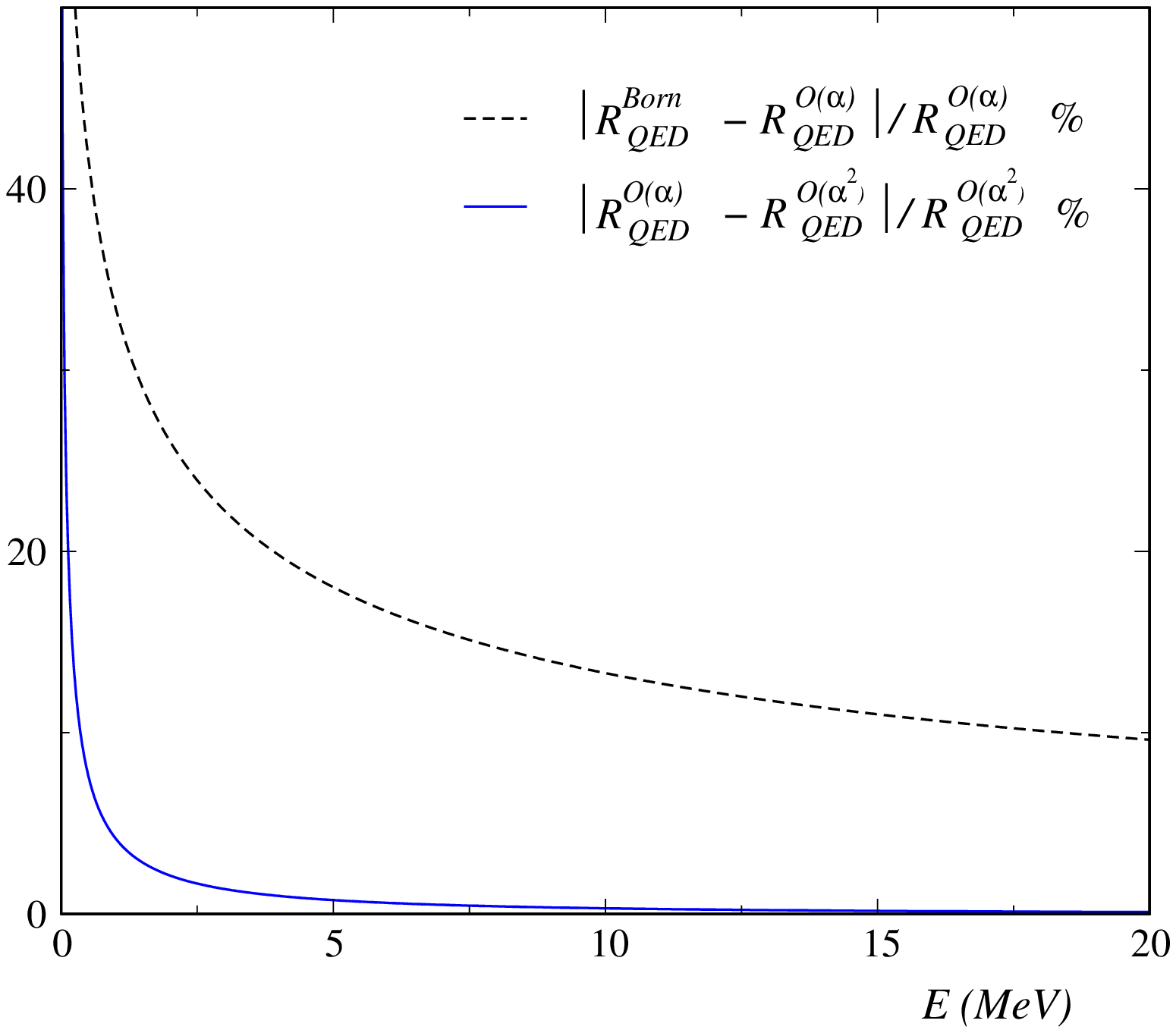}
\vspace*{0.cm}
\put(10,0){\it (a)}
\put(260,0){\it (b)}
\caption[]{\label{fig:Rsizes}\it Relative sizes of 
corrections to $R_{em}(s)$ as calculated in (a) NRQED and 
(b) QED.}  
\end{figure}

\hspace*{0.5cm}We now use formulas collected in previous sections to analyse 
the behaviour of
$\sigma(e^+e^- \to \tau^+\tau^-)$ at threshold energies. Some of the
parameters appearing in the different pieces take the following values:
\begin{itemize}
\item[$\bullet$] The $\tau$ mass, extracted from \cite{PDG}, is 
$m_{\tau}=1777.03\pm 0.30$ MeV. 
\item[$\bullet$] The two-loop running of the electromagnetic coupling constant,
defined in the $\overline{MS}$ scheme, is needed to evaluate 
$\alpha(\mu_{\mbox{\tiny $soft$}})$ and $\alpha(\mu_{\mbox{\tiny $hard$}})$, which
show up in the non-relativistic correlator and in the short-distance
coefficient, respectively. The 1- and 2-loop coefficients of the 
$\beta$-function were already given in (\ref{betacoeff}). The reference value for
the QED running coupling has been chosen by the relation
$\alpha_{\mbox{\tiny $\overline{MS}$}}(m_e^2)=\alpha$, with $\alpha=1/137.036$
the ordinary fine structure constant. 
\item[$\bullet$] The first QCD perturbative correction to the vacuum polarization of
free quarks is proportional to the strong coupling constant $\alpha_s$ (see eqs.~(\ref{Pi uds})
and (\ref{Pi cb})). At the
energy scale of $\tau$ production, it is appropriate to choose $m_{\tau}$ as 
the normalization point for $\alpha_s$; the corresponding value is
$\alpha_s(m_{\tau}^2)=0.35\pm 0.03$ \cite{PichTau}. 
\item[$\bullet$] The dependence on the various renormalization scales 
$\mu_{\mbox{\tiny $soft$}}$, $\mu_{\mbox{\tiny $hard$}}$ and 
$\mu_{\mbox{\tiny $fac$}}$ is very small. The most pronounced one
comes from variations on the scale $\mu_{\mbox{\tiny $soft$}}$ governing the
combined expansion in $\alpha$ and $v$ of the NRQED correlators. 
The logarithms of this scale over $Mv$, which show up in the
non-relativistic Green's functions, suggest taking 
$\mu_{\mbox{\tiny $soft$}} \sim Mv \sim M\alpha \simeq 13\,$MeV to minimize 
the size of the NLO and the NNLO corrections. In fact, in the range 10 MeV 
$<\mu_{\mbox{\tiny $soft$}}<$ 100 MeV the sensitivity to changes in this scale
is reduced, and we have the smallest NLO and NNLO corrections to 
$R_{\mbox{\tiny $em$}}$, varying in the whole range by less than 0.15\% and 
0.08\% respectively. The residual dependences on the other two scales 
are fully negligible.
\end{itemize}
\begin{figure}[!tb]
\begin{center}
\hspace*{-0.5cm}
\includegraphics[angle=0,width=0.9\textwidth]{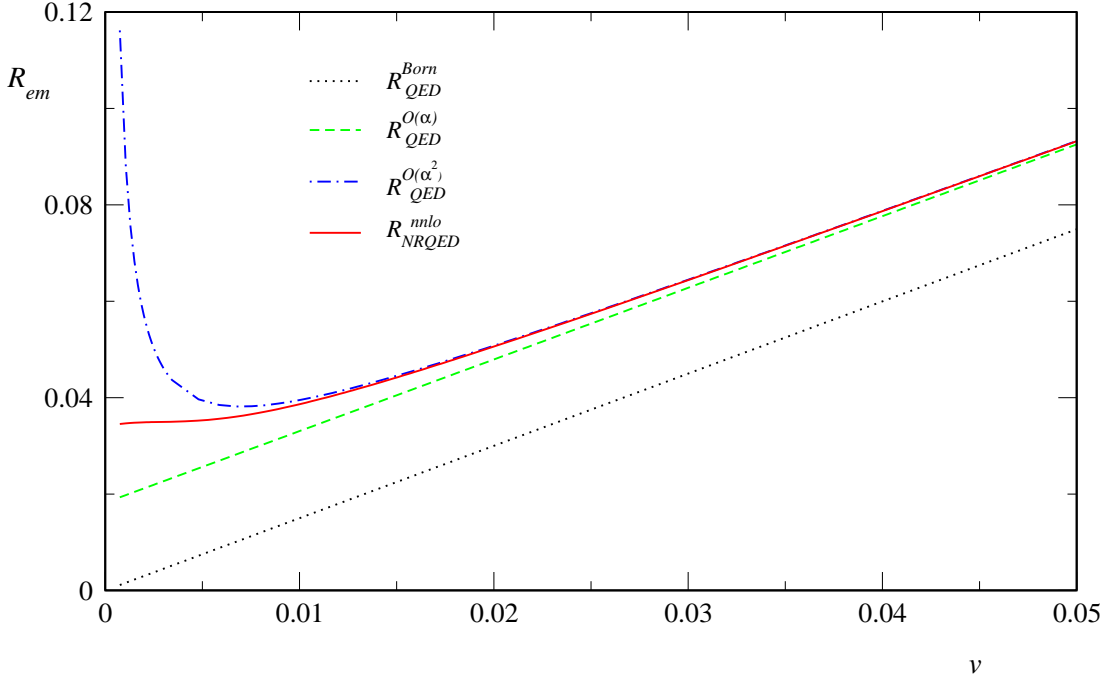}
\end{center}
\caption[]{\label{fig:Rthreshold} \it The spectral density
$R_{em}$ at low velocities in both QED and NRQED.}
\end{figure}
The need for performing resummations of the leading non-relativistic terms 
$\left( \alpha/v \right)^n [v ,v\alpha,v^2,\dots]$ is evidenced
in Figs.~\ref{fig:Rthreshold} and \ref{fig:Rsizes}. The spectral density
$R_{\mbox{\tiny $em$}}$, calculated in both
QED and NRQED, is displayed in Fig.~\ref{fig:Rthreshold} as a function 
of the $\tau$ velocity. The QED tree-level result vanishes as $v \to 0$, due to
the phase space velocity in formula (\ref{s0}), which is cancelled by the
first $v^{-1}$ term appearing in the ${\cal O}(\alpha)$ correction, making the
cross section at threshold finite. More singular terms near threshold,
$v^{-2}, v^{-3},\dots$
arising in higher-order corrections completely spoil the expected good 
convergence of the QED perturbative series in the limit $v \to 0$. 
This breakdown is clearly seen in the behaviour of the 
${\cal O}(\alpha^2)$ correction to the QED spectral density in 
Fig.~\ref{fig:Rthreshold}. This is no longer the case for the effective theory
perturbative series, whose convergence improves as we approach the threshold
point, as shown in Fig.~\ref{fig:Rsizes}a, and higher-order corrections
reduce the perturbative uncertainty inherent to any series truncated at a finite
order. In the whole energy range displayed in Fig.~\ref{fig:Rsizes}a, the
differences between the NNLO, NLO and LO results are
below 0.8\%, which indicates that the LO result, i.e. the
Sommerfeld factor, contains the relevant physics to describe the threshold
region, although NLO and NNLO corrections would be needed for more accurate 
descriptions of the total cross section.

We can safely assume that the NNLO
result for the spectral density has a theoretical uncertainty below 0.1\%
for energies close enough to threshold. At larger energies, the subleading
contributions gain importance and the convergence of the double series in
$\alpha$ and $v$ is poorer, due to the higher powers of the velocity which are
not taken into account. This is the opposite behaviour to that of the usual
perturbative QED expansion, 
Fig.~\ref{fig:Rsizes}b, where the series convergence improves as we move far
away the threshold.\par
\begin{figure}[tb!]
\includegraphics[angle=0,width=0.9\textwidth]{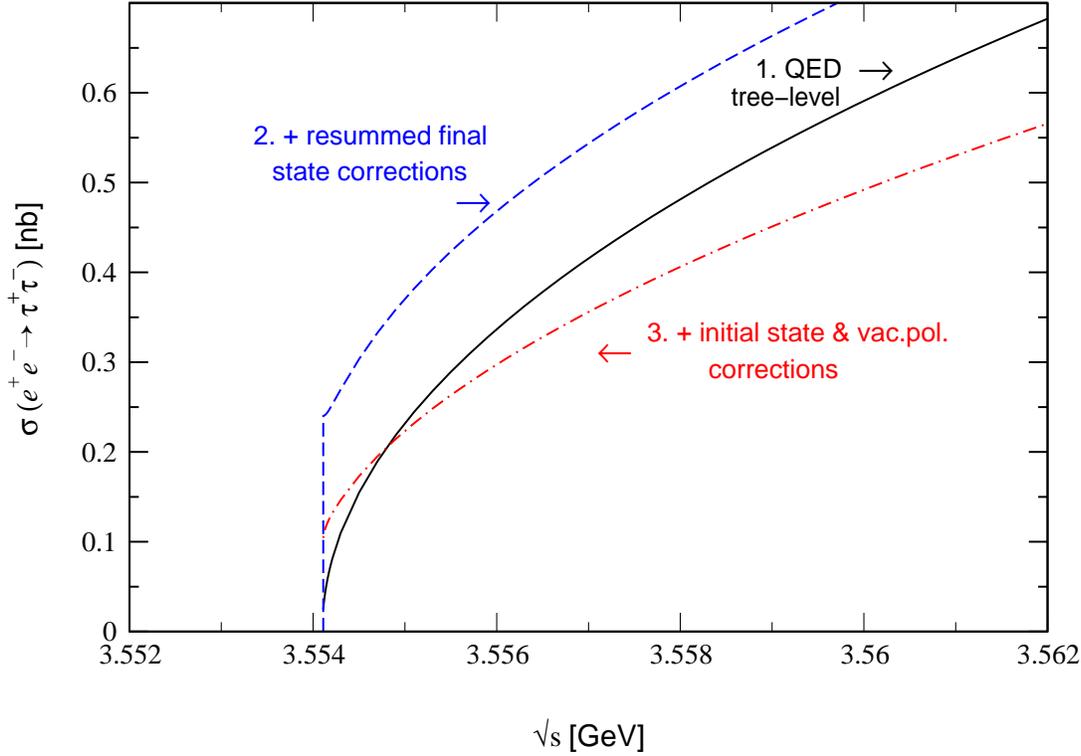}
\caption[]{\label{fig:kirk} \it The total cross section $\sigma(e^+e^-\to
\tau^+\tau^-)$ at threshold: at tree level (solid line); 
plus NNLO corrections
to final state interaction, eq.~(\ref{spectral}) (dashed line); and also 
including
radiative corrections from the initial $e^+e^-$ state and from 
vacuum polarization, eq.~(\ref{secradR}), (dash-dotted curve).}
\end{figure}
\begin{figure}[tb!]
\includegraphics[angle=0,width=0.9\textwidth]{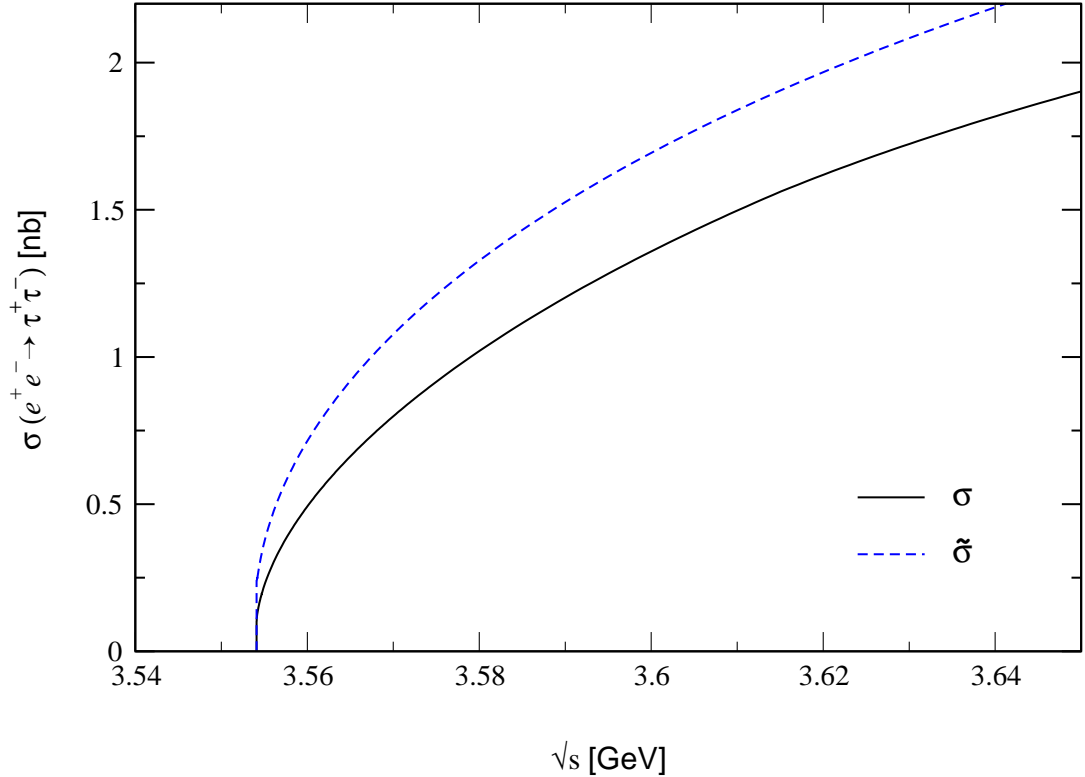}
\caption[]{\label{fig:total} \it Initial radiation effects in the total cross
section $\sigma(e^+e^-\to\tau^+\tau^-)$ up to energies around 100 MeV above 
threshold ($v\simeq 0.2$); the dashed line represents 
$\tilde{\sigma}(e^+e^-\to\tau^+\tau^-)$, which does not include 
radiative corrections from initial state, as defined in equation (\ref{spectral}).}
\end{figure}

Adding the intermediate and initial state corrections we have a complete
description of the total cross section of $\tau^+\tau^-$ production, as
shown in Fig.~\ref{fig:kirk}. Coulomb interaction between the produced $\tau$'s, governed by the parameter $\alpha/v$, becomes essential right within few
MeV above the threshold, and his effects have to be taken into account to all
orders in this parameter, making the total cross section finite in this region. 
Initial state radiation effectively reduces the available center-of-mass energy
for $\tau$ production, lowering in this way the total cross section. We can verify 
that this reduction remains at higher
energies above threshold by examining Fig.~\ref{fig:total}. A maximum energy
for the soft photons, $\Delta E=60$ MeV has been chosen to perform
the integration (\ref{secradR2}).  

We should emphasize that NNLO corrections do not modify
the predicted behaviour of the LO and NLO cross section as calculated in previous
works \cite{voloshin,perrottet}, but are essential to improve the accuracy
of experimental fits with higher precision data and, even more important, to
guarantee that the truncated perturbative series at NLO gets small
corrections from higher-order terms. In this way, we have shown that the
theoretical uncertainty of our analysis of $\sigma(e^+e^- \to \tau^+\tau^-)$ is
lower than 0.1\%, being the main sources of error our estimates of the hadronic
contribution to vacuum polarization and of the initial state radiation. The
former could be easily improved using similar techniques to those applied
to estimate $\alpha(M_Z)$, but at the energy point $\sqrt{s}=2m_{\tau}$,
including fits to $\sigma(e^+e^-\to hadrons)$ data, and the
latter, being detector dependent, should be accurately monitored and their 
effects correctly implemented in data analyses. Nevertheless, the statistical
uncertainty of the most recent experiments is still much larger than 
the theoretical one due to low statistics, 
and we should wait for future machines to improve it.
\newpage
\noindent
{\large \bf Acknowledgements}\par
\vspace{0.3cm}
\noindent 
We are grateful to J. Portol\'es for helpful discussions. We would also like to
thank M.~Eidem\"uller, M.~Jamin, M.~Perrottet and E.~de~Rafael for useful comments. 
This work has been supported by the EU TMR, EC-contract No. ERB FMRX-CT98-0169,
and by CICYT (Spain) under grant PB97-1261.
The work of P. Ruiz-Femen\'\i a has been partially supported by a FPU
scholarship of the Spanish {\it Ministerio de Educaci\'on y Cultura}.
\vfill\eject

\newcounter{erasmoA}
\renewcommand{\thesection}{\Alph{erasmoA}}
\renewcommand{\theequation}{\Alph{erasmoA}.\arabic{equation}}
\renewcommand{\thetable}{\Alph{erasmoA}}
\setcounter{erasmoA}{1}
\setcounter{equation}{0}
\setcounter{table}{0}

\section*{Appendix A}
\label{appen:R(s)NNLO QED}

\hspace*{0.5cm}The expression of the two-loop spectral density as calculated in perturbative QED and
up to NNLO in the velocity expansion ($\tilde{v}\equiv\sqrt{E/M}$) reads \cite{hoangver}:
\begin{eqnarray}
R_{\mbox{\tiny 2loop QED}}^{\mbox{\tiny NNLO}}  = 
\!\!\! &\bigg[ & \!\!\!
\frac{3}{2}\,\tilde{v}-\frac{17}{16}\,\tilde{v}^3\,+{\cal O}(\tilde{v}^4)
\,\bigg] +
\frac{\alpha(\mu_{\mbox{\tiny $h$}})}{\pi}\,\bigg[\,
\frac{3\,\pi^2}{4}-6\,\tilde{v}+\frac{\pi^2}{2}\,\tilde{v}^2\,+{\cal O}(\tilde{v}^3)
\,\bigg]
\nonumber\\[2mm]  
\!\!\!&+&\!\!\! \alpha^2(\mu_{\mbox{\tiny $h$}})\,\bigg[\,
\frac{\pi^2}{8\, \tilde{v}} 
+ \frac{3}{2}\bigg(\,
- 2 \, 
+ n_f \,\Big(\, 
    \frac{1}{6}\ln\frac{4\,\tilde{v}^2\,M^2}{\mu_{\mbox{\tiny $hard$}}^2}-\frac{5}{18}  
\,\Big)
\,\bigg) 
\nonumber\\[2mm] 
 \qquad
\!\!\!&+&\!\!\!\bigg(\, 
\frac{49\,\pi^2}{192}  
  + \frac{3}{2}\,\kappa 
  - 2\, n_f\frac{1}{\pi^2}\,  
        \, \ln\frac{M^2}{\mu_{\mbox{\tiny $hard$}}^2} 
  - \,\ln \tilde{v}
\,\bigg)\,\tilde{v}\,+{\cal O}(\tilde{v}^2)
\,\bigg]
\,.
\label{R(s)NNLO QED}
\end{eqnarray}
The constant $\kappa$ has already been defined in eq.~(\ref{kappadef}). The
renormalization point in the $\overline{MS}$ scheme has been chosen equal to
$\mu_{\mbox{\tiny $hard$}}$, and $M$ denotes the pole mass.
 
\appendix
\newcounter{erasmoB}
\renewcommand{\thesection}{\Alph{erasmoB}}
\renewcommand{\theequation}{\Alph{erasmoB}.\arabic{equation}}
\renewcommand{\thetable}{\Alph{erasmoB}}
\setcounter{erasmoB}{2}
\setcounter{equation}{0}
\setcounter{table}{0}

\section*{Appendix B}
\label{appen:NRQED}

\hspace*{0.5cm}The NRQED Lagrangian
relevant for our analysis reads
\begin{eqnarray}
\lefteqn{
{\cal{L}}_{\mbox{\tiny NRQED}} \, = \,
\frac{1}{2}\,(\,{\mbox{\boldmath $E$}}^2-
    {\mbox{\boldmath $B$}}^2\,)
+\, \psi^\dagger\,\bigg[\,
i D_t
+ c_2\,\frac{{\mbox{\boldmath $D$}}^2}{2\,M}
+ c_4\,\frac{{\mbox{\boldmath $D$}}^4}{8\,M^3}
+ \ldots}
\nonumber\\[2mm] & &
\hspace{8mm}
+  \frac{c_F\,e}{2\,M}\,{\mbox{\boldmath $\sigma$}}\cdot
    {\mbox{\boldmath $B$}}
+ \, \frac{c_D\,e}{8\,M^2}\,(\,{\mbox{\boldmath $D$}}\cdot
  {\mbox{\boldmath $E$}}-{\mbox{\boldmath $E$}}\cdot
  {\mbox{\boldmath $D$}}\,)
+ \frac{c_S\,e}{8\,M^2}\,i\,{\mbox{\boldmath $\sigma$}}\,
  (\,{\mbox{\boldmath $D$}}\times
  {\mbox{\boldmath $E$}}-{\mbox{\boldmath $E$}}\times
  {\mbox{\boldmath $D$}}\,)
+\ldots
 \,\bigg]\,\psi
\nonumber\\[2mm] & &
- \,\frac{d_1\,e^2}{4\,M^2}\,
  (\psi^\dagger{\mbox{\boldmath $\sigma$}}\sigma_2\chi^*)\,
  (\chi^T\sigma_2{\mbox{\boldmath $\sigma$}}\psi)
- \,\frac{d_2}{M^2}\,
  (\psi^\dagger\sigma_2\chi^*)\,(\chi^T\sigma_2\psi)\,
+\,
\nonumber\\[2mm] & &
+ \frac{d_3\,e^2}{6\,M^4}\,
  \Big[\,
  (\psi^\dagger{\mbox{\boldmath $\sigma$}}\sigma_2\chi^*)\,
  (\chi^T\sigma_2{\mbox{\boldmath $\sigma$}}
    (-\mbox{$\frac{i}{2}$}
  {\stackrel{\leftrightarrow}{\mbox{\boldmath $D$}}})^2\psi)
  +\mbox{h.c.}\,\Big]
+\ldots\,
\,.
\label{NRQEDLagrangian}
\end{eqnarray}
The lepton and antilepton are described by the Pauli spinors $\psi$ and $\chi$,
respectively. Antilepton bilinears and higher--order operators have been
omitted. The first line in eq.~(\ref{NRQEDLagrangian}) is related to the kinetic
term of the QED Lagrangian,  with the bilinear $\psi$ terms coming from
the expansion of the lepton relativistic energy up to ${\cal O}(1/M^3)$. Second
line terms reproduce the electromagnetic couplings of the leptons with photons
of energy lower than $M$.
Four fermion operators displayed in latter lines reproduce
production and annihilation of an $\ell^+\ell^-$ pair in a S-wave singlet
($d_2$)
or triplet ($d_1$ and $d_3$) state. Additional interaction terms between
photon fields should be introduced to simulate fermion loops. The short-distance
coefficients $c_i,d_i$ must be determined following the matching procedure up to
a certain order in $\alpha$, in order
to absorb infinities arising in calculations beyond tree level.

Which interactions are to be kept for a given precision (in $\alpha$ and $v\sim
p/M$) is dictated by counting rules. The presence of two dynamical scales in the
theory, the fermions three-momentum $\simeq Mv$, and their kinetic energies
$\simeq Mv^2$, makes the NRQED counting rules more involved than in most
effective field theories.
While the factors of $\alpha$ in a specific diagram can be read off from vertex
coefficients, powers of $v$ are also generated by internal propagators and loop
integrations. There has been a hard discussion during recent years
on how to
organize calculations within NRQED/NRQCD in a systematic expansion in $v$
\cite{NRQCDdiscussion},
especially in the context of dimensional regularization.
The situation seems to be clarified with the new
formulation proposed in Refs.~\cite{manoharvNRQED,vNRQEDrenor}.
In a cutoff scheme power counting rules for the velocity had been previously
derived by Labelle \cite{labelle} using time ordered perturbation theory
together with the Coulomb gauge to separate the ``soft'' photons (with energy
$E_\gamma \simeq Mv$) from the ``ultrasoft'' ones ($E_\gamma \simeq Mv^2$).
Although quite troublesome for calculations beyond NNLO in the velocity
expansion, these rules
give the order in $v$ of diagrams containing only soft photons
by simple dimensional analysis. Following these rules one proves that the latter
diagrams are all we need to describe low-energy interaction between the pair
of fermions up to NNLO. Moreover, soft photons have an energy
independent propagator and therefore all interactions up to NNLO can be
described in terms of potentials, being this a highly non-trivial result which
cannot be derived in the context of full QED covariant perturbation 
theory~\footnote{In terms of diagrams this statement means that only
ladder diagrams with Coulomb-like photons and contact interactions with vertex
factors up to NNLO contribute. Crossed ladder graphs vanish for soft
photons.}.

\begin{figure}[!tb]
\begin{center}
\includegraphics[angle=0,width=0.34\textwidth]{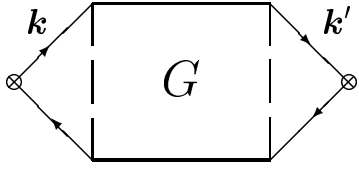}
\end{center}
\caption[]{\label{fig:Gevol} \it Graphical representation of the NRQED
vector-current correlator diagrams: the lepton pair $\ell^+\ell^-$ is
created and annihilated by the coupling
${\psi}^\dagger {\mbox{\boldmath $\sigma$}} \, \chi$ in (\ref{A1}),
and all the intermediate diagrams of the
$\ell^+\ell^-$ non-relativistic NNLO interaction
are resummed in the Green's function $G(E)$.}
\end{figure}

The effective
$\gamma \tau^+\tau^-$ coupling seen by the non-relativistic leptons is given
by the expansion of the QED current in terms of the operators of the low-energy
theory:
\begin{equation}
j^{k}_{\mbox{\tiny NR}}(x)  =  b_1\,\Big(\psi^{\dagger} \sigma^k
\chi\Big)(x) -
\frac{b_2}{6 M^2}\,\Big({\psi}^\dagger \sigma^k
(\mbox{$-\frac{i}{2}$}
\stackrel{\leftrightarrow}{\mbox{\boldmath $D$}})^2
 \chi\Big)(x) + \ldots
\,.
\label{corrienteNR}
\end{equation}
We have only quoted the terms which are needed at NNLO. The first piece is a
dimension-three current while the second has dimension-five and it is already of
NNLO, as dictated by counting rules \cite{labelle} due to the presence of the
$1/M^2$ factor. Notice that both pieces have quantum numbers $^3S_1$. There is
another dimension five current, describing $^3D_1$ $\tau^+\tau^-$ production
which however would not contribute to the NNLO cross section because the
correlator of the product of a $^3S_1$ current and a $^3D_1$ one vanishes.   
The Wilson coefficients of the NRQED $^3S_1$ current encode the effects of the
hard modes which have been integrated out. The coefficient $b_1$ needs 
to be known at order
$\alpha^2$, while $b_2=1$ at NNLO. Inserting expansion (\ref{corrienteNR}) into
the correlation function (\ref{Rcorrelator}) leads to the NRQED expression of
the ratio $R_{\mbox{\tiny $em$}}$ at NNLO
\begin{eqnarray}
R^{\mbox{\tiny NNLO}}_{\mbox{\tiny $em$}}(q^2)  = 
\frac{4\,\pi}{q^2} \, \mbox{Im} \Big( C_1\,
\Big[\, {\cal{A}}_1 (E)
\,\Big] -\frac{1}{6M^2} C_2\, \Big[\,
{\cal{A}}_2 (E) \,\Big] \Big)
\,,
\label{R(s)NNLO}
\end{eqnarray}
where
\begin{eqnarray}
{\cal{A}}_1 & = & -i \int d^4x \, e^{iqx}\,
\Big\langle \, 0 \, \Big| T
\, \Big({\psi}^\dagger {\mbox{\boldmath $\sigma$}} \, \chi
\cdot {\chi}^\dagger {\mbox{\boldmath $\sigma$}} \,
\psi \Big)\,\Big| \, 0 \, \Big\rangle
\,,
\label{A1}
\\[2mm]
{\cal{A}}_2 & = & -i \int d^4x \, e^{iqx}\,
\Big\langle \, 0 \, \Big| T
\, \Big({\psi}^\dagger {\mbox{\boldmath $\sigma$}} \, \chi
\cdot {\chi}^\dagger  {\mbox{\boldmath $\sigma$}} \,
(\mbox{$-\frac{i}{2}$}\stackrel{\leftrightarrow}{\mbox{\boldmath $D$}})^2
\psi + \mbox {h.c.} \Big)\,
\Big| \, 0 \, \Big\rangle
\,.
\label{A2}
\end{eqnarray}   
The short distance coefficients read $C_1=(b_1)^2$ and $C_2=1$. The correlators
${\cal{A}}_1$ and ${\cal{A}}_2$ contain the non-relativistic interactions
derived from the NRQED Lagrangian. Such interactions, at NNLO, are purely
described by instantaneous potentials, similar to those used in
familiar quantum mechanics. Therefore, once the lepton pair
is created by the external current with relative momentum
${\mbox{\boldmath $k$}}$ and until it is
annihilated,
the four point function describing their evolution reduces to a
Schr\"odinger Green function for a two-body system
with kinetic energy
$E=\sqrt{s}-2M$, see Fig.~\ref{fig:Gevol}.  
The exact relation for ${\cal{A}}_1$ reads
\begin{eqnarray}
{\cal{A}}_1(E) & = & \mbox{Tr}
\int\frac{d^3 \mbox{\boldmath $k$}}{(2\pi)^3}
\int\frac{d^3 \mbox{\boldmath $k$}^\prime}{(2\pi)^3}\,
\mbox{\boldmath $\sigma$}\,
\tilde{G}({\mbox{\boldmath $k$}},{\mbox{\boldmath $k$}^\prime};E)\,
\mbox{\boldmath $\sigma$}
\nonumber\\[2mm] & = &
6\,\Bigl[ \,\lim_{r,r^\prime \to 0}
 \, G({\mbox{\boldmath $r$}},{\mbox{\boldmath $r$}^\prime};E)\,\Bigr]
\,,
\label{correlador1Green}
\end{eqnarray}
where we have used the identity
$\mbox{Tr}(\mbox{\boldmath $\sigma$}\cdot\mbox{\boldmath $\sigma$})=
3\,\mbox{Tr}(\mbox{I})=6$. One can check that eq.~(\ref{correlador1Green})
gives the right proportionality factor between ${\cal{A}}_1$ and
$G$ just considering the free case. There is no extra factor coming
from the different normalizations of the relativistic and the non-relativistic
quantities.

For the ${\cal{A}}_2$ correlator we have
\begin{eqnarray}
{\cal{A}}_2(E) & = & \mbox{Tr}
\int\frac{d^3 \mbox{\boldmath $k$}}{(2\pi)^3}
\int\frac{d^3 \mbox{\boldmath $k$}^\prime}{(2\pi)^3}\,
({\mbox{\boldmath $k$}}^2+{\mbox{\boldmath $k$}^\prime}^2)\,
\tilde{G_c}({\mbox{\boldmath $k$}},{\mbox{\boldmath $k$}^\prime};E)
\,
\nonumber\\[2mm] & = &-6
\left( {\mbox{\boldmath $\nabla$}}^{2}_{r}+
{\mbox{\boldmath $\nabla$}}^{2}_{r^\prime}\right)
G_c({\mbox{\boldmath $r$}},{\mbox{\boldmath $r$}^\prime};E)|_{r,r^\prime\to 0}
\,.
\label{correlador2Green}
\end{eqnarray}
As ${\cal{A}}_2$ is already of NNLO, only the Green's function for the Coulomb
potential shall be considered. Relation (\ref{correlador2Green}) can be further
simplified by using the Schr\"odinger eq.~(\ref{Schrodingerfull}),
retaining just the
LO piece of $V_c(r)$. For the imaginary part, we have
\begin{equation}
 -\frac{{\mbox{\boldmath $\nabla$}}^2_r}{M}
\,\mbox{Im}\, G_c({\mbox{\boldmath $r$}},{\mbox{\boldmath $r$}^\prime};E)\,=\,
\left(E-V_{c}^{\mbox{\tiny LO}}(r)
\,\right)\,\mbox{Im}\, G_c({\mbox{\boldmath $r$}},{\mbox{\boldmath $r$}^\prime};E)
\, = \,
\left(E+\frac{\alpha}{r}\,\right)\,
\mbox{Im}\, G_c({\mbox{\boldmath $r$}},{\mbox{\boldmath $r$}^\prime};E)
\,.
\label{SchrodingerIm}
\end{equation}
In the limit $r,r^\prime\to 0$, the term $\alpha/r\,\mbox{Im}G_c$ represents an
ultraviolet divergence which must be regularized. Following the direct matching
procedure \cite{hoangvac} to fix the value of the short distance coefficient
$C_1$ allows us to drop power-like divergences, such as
$\alpha/r\,\mbox{Im}G_c|_{r\to 0}$, which must cancel with similar ultraviolet
divergences in $C_1$ in the final expression for the total cross section.
Therefore we can safely substitute $\mbox{Im}\,{\cal{A}}_2$ by
$12EM
 \, G_c({\mbox{\boldmath $r$}},{\mbox{\boldmath $r$}^\prime};E)
 |_{r,r^\prime\to 0}$ in 
(\ref{R(s)NNLO}) to get the complete relation between the spectral density
at NNLO and the non-relativistic Green's functions:
\begin{eqnarray}
R^{\mbox{\tiny NNLO}}_{\mbox{\tiny $em$}}(q^2)  = 
\frac{6\,\pi}{M^2} \, \mbox{Im} \Big( C_1\,
G({\mbox{\boldmath $0$}},{\mbox{\boldmath $0$}};E)
\,-\frac{4E}{3M} \,
G_c({\mbox{\boldmath $0$}},{\mbox{\boldmath $0$}};E)
\Big)
\,,
\label{R(s)NNLO G}
\end{eqnarray}
where we have expanded the relation $q^2=(2M+E)^2$ to first
order in $E/2M$, which is already a NNLO contribution.

\appendix
\newcounter{erasmoC}
\renewcommand{\thesection}{\Alph{erasmoC}}
\renewcommand{\theequation}{\Alph{erasmoC}.\arabic{equation}}
\renewcommand{\thetable}{\Alph{erasmoC}}
\setcounter{erasmoC}{3}
\setcounter{equation}{0}
\setcounter{table}{0}

\section*{Appendix C}
\label{appen:Greens}

\hspace*{0.5cm}The well-known
Coulomb Green's function \cite{schwinger}, solution of the LO Hamiltonian,
at the origin reads ($\tilde{v}\equiv\sqrt{E/M}\,$)
\begin{equation}
G_c^r({\mbox{\boldmath $0$}},{\mbox{\boldmath $0$}};E)
 \, =  \,
\frac{M^2}{4\pi}\, \bigg(\, 
  i\,\tilde{v}  - \alpha(\mu_s) \bigg[{\mbox{ln}}\,
  \Big(-i\frac{M\tilde{v}}{\mu_{\mbox{\tiny $fac$}}}\Big)+\gamma+
  {\Psi}\Big(1 - i\,\frac{\alpha(\mu_s)}{2\,\tilde{v}}\Big) \,\bigg] \bigg)
\,,
\label{coulombGreen-renor}
\end{equation}
where $\Psi(z)=\frac{d}{dz}\log \Gamma(z)$ and $\Gamma(z)$ is the Euler
Gamma function. The superscript `r' stands for `renormalized', since in the
short distance limit $r,r^\prime \to 0$ the Coulomb Green's function, and some
of the $\delta G$, have $1/r$ and $\log (r)$ divergent terms.
Following the lines of previous papers \cite{hoangteubner,melnikov}
power-like divergences are subtracted and ultraviolet log-terms are
regularized by introducing a cutoff $\mu_f$ and hence subtracting
the energy-independent part. However, the imaginary part
of $G_c$ has no ultraviolet divergent terms, so they would not
contribute to the total cross section. This is not longer the case for the
corrections $\delta_{\mbox{\tiny Ki,BF}}G$ and $\delta_{\mbox{\tiny An}}G$,
and their (imaginary part) residual dependence
on the $\mu_f$-scale will be canceled with the scale dependence of the
coefficient $C_1$, which is
determined using the ``direct matching procedure'' \cite{hoangvac} described
at the end of section~\ref{sec:NRQED}. We quote the result for $\delta_{\mbox{\tiny Ki,BF}}G$
($C_F\to 1, T_F\to 1$ and $C_A\to 0$ for the U(1) group)
\cite{hoangteubner}:
\begin{eqnarray}
\delta_{\mbox{\tiny Ki,BF}}G(0,0;E)&=&
\frac{\alpha(\mu_s) M^2}{4\pi}\,\Bigg(i\,
{5\over 8}\,\frac{\tilde{v}^3}{\alpha(\mu_s)}-
2\tilde{v}^2 \left[
\ln \left(-i\frac{M\tilde{v}}{\mu_{\mbox{\tiny $fac$}}}\right)+\gamma
+\Psi \left(1-i
\frac{\alpha(\mu_s)}{2\tilde{v}} \right) \right]
\nonumber\\[2mm]
&&+\,\,\,i\,{11\over 16}\alpha(\mu_s) \tilde{v}\,
\Psi^{\prime}\left(1-i {\alpha(\mu_s) \over 2\tilde{v}}\,\right) \Bigg)
+{4\pi\over 3}{\alpha(\mu_s)\over M^2}G_c^r(0,0,E)^2
\,.
\label{GBreitKin}
\end{eqnarray}

The integration for the $V_{\mbox{\tiny An}}$ potential is trivial, and the
resulting (renormalized) correction $\delta_{\mbox{\tiny An}}G$ reads
\begin{eqnarray}
\delta_{\mbox{\tiny An}}G(0,0;E)=
-2{\alpha(\mu_s)\pi\over M^2}G_c^r(0,0,E)^2
\,.
\label{GAni}
\end{eqnarray}

The ${\cal O}(\alpha)$ correction to the Coulomb potential,
$V_{c}^{\tiny (1)}(r)$, must be iterated twice because it is a NLO
contribution. The corresponding corrections 
$\delta^{\mbox{\tiny NLO}}_{\tiny 1} G$ and
$\delta^{\mbox{\tiny NNLO}}_{\tiny 1} G$  have been calculated in
\cite{kuhn} and \cite{peninphyslett}, respectively. The
details of their calculation can be found therein. Their final expressions read:
$$
\delta^{\mbox{\tiny NLO}}_1G(0,0;E)=\bigg(\frac{\alpha(\mu_s)}{4\pi}\bigg)^2
M^2\left\{
\sum_{m=0}^\infty F^2(m)(m+1)
\left(C_0^1+(L(v)+\Psi(m+2))C_1^1\right)
\right.
$$
\[
-2\sum_{m=1}^\infty\sum_{n=0}^{m-1}
F(m)F(n)
{n+1\over m-n}C_1^1 
+2\sum_{m=0}^\infty F(m)
\left(C_0^1+(L(v) -
2\gamma-\Psi(m+1))C_1^1\right)
\]
\begin{equation}
\left.+L(v)C_0^1+\left(-\gamma L(v)+
{1\over 2}L(v)^2\right)C_1^1
\right\}\,,
\label{G1NLO}
\end{equation}
and
 \[
\delta^{\mbox{\tiny NNLO}}_1G(0,0;E)=i\left({\alpha(\mu_s)\over 4\pi}\right)^2
{\alpha(\mu_s)^2\over 4\pi}{M^2\over 2v}
\left\{ \sum_{m=0}^\infty H^3(m)(m+1)
\right.\cdot
\]
$$
\cdot\left(C_0^1+
\left(\Psi(m+2)+
L(v)\right)C_1^1\right)^2
$$
\[
-2\sum_{m=1}^\infty\sum_{n=0}^{m-1}{n+1\over m-n}C_1^1
\left(H^2(m)H(n)\left(C_0^1+\left(\Psi(m+2)+
L(v)-{1\over 2}{1\over m-n}\right)C_1^1\right)
\right.
\]
\begin{equation}
\left.
+H(m)H^2(n)\left(C_0^1+\left(\Psi(n+2)+
L(v)-{1\over 2}{n+1\over (m-n)(m+1)}\right)C_1^1\right)\right)
\label{G1NNLO}
\end{equation}
\[
+2(C_1^1)^2\left(\sum_{m=2}^\infty\sum_{l=1}^{m-1}\sum_{n=0}^{l-1}
{H(m)H(n)H(l)}{n+1\over (l-n) (m-n)}\right.
\]
$$
+\sum_{m=2}^\infty\sum_{n=1}^{m-1}\sum_{l=0}^{n-1}
{H(m)H(n)H(l)}{l+1\over (n-l)(m-n)}
$$
\[
\left.\left.
+\sum_{n=2}^\infty\sum_{m=1}^{n-1}\sum_{l=0}^{m-1}
{H(m)H(n)H(l)}{(l+1)(m+1)\over (n+1)(n-l)(n-m)}\right)\right\}\,,
\]
with
\begin{equation}
F(m)=\frac{i}{2(m+1)}\,\frac{\alpha(\mu_s)}{v}\,
\bigg(m+1-i\,\frac{\alpha(\mu_s)}{2v}\bigg)^{-1}
\,,
\label{F(m)}
\end{equation}
\begin{equation}
L(v)=-{\mbox{ln}}\bigg(-2i\frac{Mv}{\mu_s}\bigg)
\,,
\label{L(v)}
\end{equation}
and finally 
$$
H(m)=\left(m+1-i{\displaystyle {\alpha(\mu_s) \over 2v}}\right)^{-1}. 
$$ 
The constants $C_0^1$, $C_1^1$ are defined in terms of $\beta_1$ (\ref{beta-def}):
\begin{eqnarray}
C_0^1 & = & a_1+2\beta_1\gamma
\,,
\nonumber\\[2mm]
C_1^1 & = & 2\beta_1
\,.
\label{C01,C11}
\end{eqnarray}
The iteration of the ${\cal O}(\alpha^2)$ piece, 
$V_{c}^{\tiny (2)}(r)$, was also computed in \cite{kuhn}:
\[
\delta_2G(0,0;E)=\left({\alpha(\mu_s)\over 4\pi}\right)^2{\alpha(\mu_s) M^2\over
4\pi}\left\{ \sum_{m=0}^\infty F^2(m)
\left((m+1)\left(C_0^2+L(v) C_1^2+L^2(k) C_2^2\right)
\right.\right.
\]
$$
\left.
+(m+1)\Psi(m+2)\left(C_1^2+2L(v)
C_2^2\right)+I(m)C_2^2\right) 
$$
\begin{equation}
+2\sum_{m=1}^\infty\sum_{n=0}^{m-1}
F(m)F(n)\left(-
{n+1\over m-n}\left(C_1^2 +2L(v) C_2^2\right)
+J(m,n)C_2^2\right)
\label{G22}
\end{equation}
\[
+2\sum_{m=0}^\infty F(m)
\left(C_0^2+L(v) C_1^2+
(L^2(v)+K(m))C_2^2-(2\gamma+
\Psi(m+1))
\left(C_1^2+2L(v) C_2^2\right)\right)
\]
$$
\left.+L(v)C_0^2+\left(-\gamma L(v)+
{1\over 2}L^2(v)\right)
C_1^2 
+N(v)C_2^2\right\}\,,
$$
with the functions $I(m),J(m,n),K(m),N(v)$ defined as
\[
I(m)=(m+1)\left(\Psi^2(m+2)-\Psi^{\prime}(m+2)+{\pi^2\over3}-
{2\over(m+1)^2}\right)
\]
$$
-2(\Psi(m+1)+\gamma),
$$
$$
J(m,n)= 2{n+1\over m-n}\left(\Psi_1(m-n)-{1\over n+1}+2\gamma\right)
$$
\[
+2{m+1\over m-n}(\Psi(m-n+1)-\Psi(m+1)),
\]
$$
K(m)=2(\Psi(m+1)+\gamma)^2+\Psi^{\prime}(m+1)-\Psi^2(m+1)+2\gamma^2,
$$
\[
N(v)=\left(\gamma+{\pi^2\over 6}\right)L(v)
-\gamma L^2(v)+{1\over 3}
L^3(v)\,,
\]
and the constants
$$
C_0^2=\Big(\frac{\pi^2}{3}+4\gamma^2\Big)+\beta_1^2+2(\beta_2+2\beta_1
a_1)\gamma+a_2\,,
$$
$$
C_1^2=2(\beta_2+2\beta_1 a_1)+8\beta_1^2\gamma\,,
$$
\[
C_2^2=4\beta_1^2\,.
\]

None of the above mentioned Coulomb $\delta G$ corrections have energy dependent
ultraviolet terms on their imaginary part, so no matching is necessary for them.


\begin{thebibliography}{99}

\bibitem{KI:87}
   J. Kirkby, {\it A $\tau$--Charm Factory at CERN}, CERN-EP/87-210 (1987).
  
\bibitem{JO:87} 
   J.M. Jowett, {\it Initial Design of a $\tau$--Charm Factory at CERN}
   CERN LEP-TH/87-56 (1987); {\it The $\tau$--Charm Factory Storage Ring},
   CERN LEP-TH/88-22 (1988).

\bibitem{tcfSLAC}
  Proc. Tau--Charm Factory Workshop (SLAC, California, 1989), ed.
  L.V. Beers, SLAC-Report-343 (1989).
 
\bibitem{marbella}
  Proc. 3rd Workshop on the Tau-Charm Factory (Marbella, Spain, 1993),
  eds. J.~Kirkby and R.~Kirkby (Editions Fronti\`eres, Gif-sur-Yvette,
  1994).

\bibitem{cornell} Workshop on Prospects for CLEO/CESR with 
$3<E_{\mbox{\tiny CM}}<5$~GeV, Cornell, May 2001.

\bibitem{PI:94}
  A. Pich, {\it Perspectives on Tau--Charm Factory Physics},
  in \cite{marbella} p.~767.

\bibitem{BES}
  J.Z. Bai {\it et al} (BES), {\it Phys. Rev.} {\bf D53} (1996) 20.

\bibitem{voloshin} M.B. Voloshin, {\it Topics in Tau Physics at a
 Tau-Charm Factory}, TPI-MINN-89/33-T (1989), unpublished
 [hep-ph/9312358].

\bibitem{perrottet} M. Perrottet, {\em An Improved Calculation of
  $\sigma (e^+\,e^- \to \tau^+ \,\tau^-)$ Near Threshold},
  in \cite{marbella} p.~89.

\bibitem{SV:94} B.H. Smith and M.B. Voloshin, {\it Phys. Lett.} {\bf B324}
  (1994) 117.
  
\bibitem{lepage} W. E. Caswell and G. P. Lepage, {\em Phys. Lett.} {\bf B167}
(1986) 437.

\bibitem{jamin} M. Jamin and A. Pich, {\em Nucl. Phys.} {\bf B507} (1997) 334.

\bibitem{kuhn} J. H. K\"uhn, A. A. Penin and A. A. Pivovarov, {\em Nucl. Phys.}
{\bf B534} (1998) 356.

\bibitem{bb} 
K. Melnikov and A. Yelkhovsky, {\em Phys. Rev.} {\bf D59} (1999) 114009; \\
A.A. Penin and A.A. Pivovarov, {\em Nucl. Phys.} {\bf B549} (1999) 217;\\
M. Beneke and A. Signer, {\em Phys. Lett.} {\bf B471} (1999) 233;\\
A.H. Hoang, {\em Phys. Rev.} {\bf D59} (1999) 014039.

\bibitem{topprodsummary}
A.H. Hoang {\it et al.},
{\em Eur.\ Phys.\ J.\ direct} {\bf C3}, (2000) 1.

\bibitem{barbi} R. Barbieri, J.A. Mignaco and E. Remiddi, {\em Nuovo Cimento}
{\bf 11A} (1972) 824.

\bibitem{kuraev} E.A. Kuraev and V.S. Fadin, {\em Yad. Fiz.} {\bf 41} (1985) 733 [{\em Sov. J. Nucl.
Phys.} {\bf41} (1985) 466].

\bibitem{hoangver} A.H. Hoang, {\em Phys. Rev.} {\bf D56} (1997) 7276.

\bibitem{sommer} A. Sommerfeld, {\em Atombau und Spektrallinien}, Vol.II,
Vieweg, Braunschweig, 1939.

\bibitem{labelle} P. Labelle, {\em Phys. Rev.} {\bf D58} (1998) 093913.

\bibitem{berends} F.A. Berends, K.J.F. Gaemers and R. Gastmans, {\em Nucl. Phys.} {\bf B57} (1973)
381.

\bibitem{hoangteubner} A. H. Hoang and T. Teubner, {\em Phys. Rev.} {\bf D58}
(1998) 114023.

\bibitem{fischler&billoire} W. Fischler, {\em Nucl. Phys.} {\bf B129} (1977)
157 ; \\
A. Billoire, {\em Phys. Lett.} {\bf B92} (1980) 343.

\bibitem{peter} M. Peter, {\em Nucl. Phys.} {\bf B501} (1997) 471;\\
Y. Schr\"oder, {\em Phys. Lett.} {\bf B447} (1999) 321.

\bibitem{landau} L. D. Landau and E. M. Lifshitz, {\em Relativistic Quantum
Theory, Part 1} (Pergamos, Oxford, 1974)

\bibitem{peninphyslett} A. A. Penin and A. A. Pivovarov, {\em Phys. Lett.}
{\bf B435} (1998) 413.

\bibitem{hoangvac} A.H. Hoang, {\em Phys. Rev.} {\bf D57} (1998) 1615.

\bibitem{kallen-sabry} G. K\"allen and A.
Sabry, Kgl. Danske Videnskab. Selskab Mat. Fys. Medd. {\bf 29} (1955) No.17.


\bibitem{guerrero} F. Guerrero and A. Pich, {\em Phys. Lett.} {\bf B412}
(1997) 382.

\bibitem{RChiT} 
G.Ecker, J. Gasser, A. Pich and E. de Rafael, {\em Nucl. Phys.} {\bf B321}
(1989) 311; \\
G.Ecker, J. Gasser, H. Leutwyler, A. Pich and E. de Rafael, {\em Phys. Lett.} 
{\bf B223} (1989) 425.

\bibitem{jorge} D. G\'omez-Dumm, A. Pich and J. Portol\'es,
{\em Phys. Rev.} {\bf D62} (2000) 054014.

\bibitem{jegerlehner} S. Eidelman and F. Jegerlehner, {\em Z. Phys.} {\bf C67}
(1995) 585.

\bibitem{davier} M. Davier and A. H\"ocker, {\em Phys. Lett.} {B435} (1998) 427.

\bibitem{Portoles} A. Pich and J. Portol\'es, {\em The vector form-factor of the
pion from unitariry and analyticity: a model independent approach}, 
hep-ph/0101194.

\bibitem{penin2} A.A. Penin and A.A. Pivovarov, {\em Analytical results for
$e^+e^-\to \bar{t}t$ and $\gamma \gamma \to \bar{t} t$ observables near the
threshold up to the next-to-next-to leading order of NRQCD}, hep-ph/9904278.

\bibitem{hoang-teubner2} A.H. Hoang and T. Teubner,
{\em  Phys. Rev.} {\bf D60} (1999) 114027.

\bibitem{PDG}
Particle Data Group, {\it Review of Particle Properties},
{\em Eur.\ Phys.\ J.\ } {\bf C15} (2000) 1. 

\bibitem{PichTau} A. Pich,{\em Tau Physics: Theoretical Perspective},
hep-ph/0012297; \\
{\em Tau Physics}, Proc. 19th Int. Symposium on Lepton and Photon Interactions
at High Energies (Stanford, 1999), eds. J. Jaros and M. Peskin (World
Scientific, Singapore, 2000) 157.

\bibitem{NRQCDdiscussion} M.E. Luke and A.V. Manohar, {\em Phys. Rev.} {\bf D55}
(1997) 4129;\\
A.V. Manohar, {\em Phys. Rev.} {\bf D56} (1997) 230;\\
A. Pineda and J. Soto, {\em Nucl. Phys. Proc. Suppl.} {\bf 64} (1998) 428;\\
M. Beneke and V.A. Smirnov, {\em Nucl. Phys.} {\bf B522} (1998) 321;\\
H.W. Griesshammer, {\em Phys. Rev.} {\bf D58} (1998) 094027.

\bibitem{manoharvNRQED} M.E. Luke, A.V. Manohar and I.Z. Rothstein,
{\em  Phys. Rev.} {\bf D61} (2000) 074025.

\bibitem{vNRQEDrenor} A.V. Manohar, J. Soto, I.W. Stewart, {\em Phys.Lett.}
{\bf B486} (2000) 400.

\bibitem{schwinger} E.H. Wichmann and C. H. Woo, {\em J. Math. Phys.} {\bf 2}
(1961) 178; \\
L. Hostler, {\em J. Math. Phys.} {\bf 5} (1964) 591; \\
J. Schwinger, {\em J. Math. Phys.} {\bf 5} (1964) 1606.

\bibitem{melnikov} K. Melnikov and A. Yelkhovsky, {\em Nucl. Phys.} {\bf B528}
(1998) 59.

\end{thebibliography}
\end{document}